%% file: 000-Main.tex
\documentclass[compsoc, conference, a4paper, 10pt, times]{IEEEtran}

\usepackage{cite}
\usepackage{amsmath,amssymb,amsfonts}
\usepackage{algorithm}
\usepackage{algpseudocode}
\usepackage{graphicx}
\usepackage{textcomp}
\usepackage{xcolor}
\usepackage{booktabs}
\usepackage[hidelinks]{hyperref}
\usepackage{tikz}
\usepackage{enumitem}
\usepackage{float}
\usepackage{subcaption}
\usepackage{dblfloatfix}
\usepackage[normalem]{ulem}
\useunder{\uline}{\ul}{}

\begin{document}

\title{Multi-Factor Credential Hashing for Asymmetric Brute-Force Attack Resistance}


\author{\IEEEauthorblockN{Vivek Nair}
\IEEEauthorblockA{\textit{UC Berkeley} \\
vcn@berkeley.edu
}
\and
\IEEEauthorblockN{Dawn Song}
\IEEEauthorblockA{\textit{UC Berkeley} \\
dawnsong@berkeley.edu
}
}

\maketitle

\begin{abstract}

Since the introduction of bcrypt in 1999, adaptive password hashing functions, whereby brute-force resistance increases symmetrically with computational difficulty for legitimate users, have been our most powerful post-breach countermeasure against credential disclosure. Unfortunately, the relatively low tolerance of users to added latency places an upper bound on the deployment of this technique in most applications. In this paper, we present a multi-factor credential hashing function (MFCHF) that incorporates the additional entropy of multi-factor authentication into password hashes to provide asymmetric resistance to brute-force attacks. MFCHF provides full backward compatibility with existing authentication software (e.g., Google Authenticator) and hardware (e.g., YubiKeys), with support for common usability features like factor recovery.
The result is a $10^6$ to $10^{48}$ times increase in the difficulty of cracking hashed credentials, with little added latency or usability impact.

\end{abstract}

\input{010-Introduction}
\input{020-Background}

\begin{figure*}[!b]
\centering
\includegraphics[width=\linewidth]{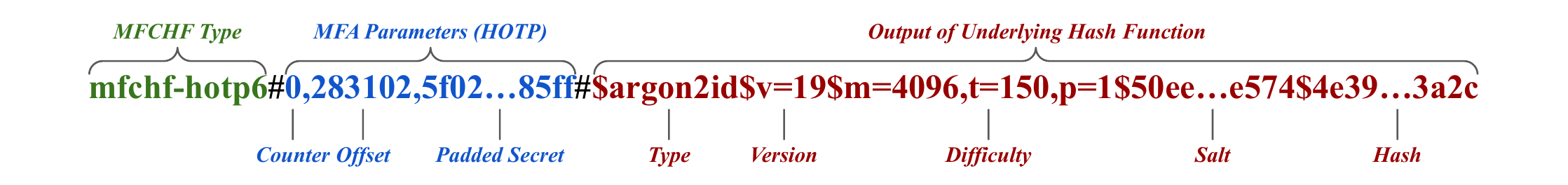}
\caption{\label{fig:hotp6}Example hash output generated by MFCHF with password and HOTP factors (mfchf-hotp6).}
\end{figure*}

\noindent 

\input{030-Problem-Statement}
\input{040-MFCHF}
\input{050-Features}

\input{090-Demo}
\input{100-Evaluation}

\input{150-Security-Analysis}

\input{200-Related-Work}
\input{210-Future-Work}
\input{220-Conclusion}

\section*{Availability}

\noindent Our GitHub repository, which contains the source code and data used to produce the results presented in this paper, is available here: \url{https://github.com/mfchf/paper}

\section*{Acknowledgments}
\noindent This work was supported in part by the National Science Foundation, the National Physical Science Consortium, the Fannie and John Hertz Foundation, and the Berkeley Center for Responsible, Decentralized Intelligence. Any opinions, findings, and conclusions or recommendations expressed in this material are those of the authors and do not necessarily reflect the views of the supporting entities.

\bibliographystyle{plainurl}
\bibliography{900-References}

\appendices

\input{950-Appendix-Demo}
\input{910-Appendix-Specs}
\input{920-Appendix-Table}
\input{930-Appendix-Results}
\input{940-Appendix-Algorithms}

\end{document}

%% file: 010-Introduction.tex
\section{Introduction}

Despite unprecedented levels of information security spending \cite{gartner_spending}, the frequency and severity of data breaches continue to experience exponential yearly growth \cite{cost_of_cybercrime}. While countless factors are at play, stolen or compromised credentials are both the largest individual \textit{cause} of data breaches, accounting for over 80\% of breaches on record \cite{ibm_cost_2022, dbir}, and the largest \textit{result} of data breaches, being present in nearly 80\% of stolen databases \cite{privacyrights, hibp}.
Thus, data breaches effectively form a negative feedback loop, whereby stolen credentials from one data breach are often used to compromise several further systems.
Preventing the extraction of plaintext credentials from stolen data is an important objective toward breaking this cycle.

Salted hashing of stored passwords has long been our strongest tool for preventing the disclosure of credentials after a data breach has occurred.
Unfortunately, the low complexity of typical passwords \cite{florencio_large_2006} and the extreme hash rate of modern ASICs \cite{bitmain} make password hashes stored using standard cryptographic hash functions highly susceptible to brute-force attacks. However, ``adaptive'' hash functions like bcrypt \cite{bcrypt} provide a convenient solution to this problem by allowing their computational complexity to increase over time in accordance with improvements in overall computational power.
While more hardware-resistant hash functions like Argon2 \cite{argon2} have since been introduced, the paradigm of adding artificial computational difficulty as the primary means of strengthening the resistance of password hashes has remained relatively unchanged since the introduction of bcrypt in 1999.

Today, the use of adaptive password hashing has proven necessary but insufficient to prevent password disclosure following data breaches, as most users' low tolerance for added latency \cite{arapakis_impact_2021} has effectively placed an upper bound on the extent to which such techniques can be utilized. As a result, some recent data breaches have seen credential disclosure ratios of nearly 50\% despite using strong adaptive password hashes (see \S\ref{sec:ratio}).

Unsurprisingly, coinciding with this alarming trend in the rate of password disclosure has been the widespread adoption of multi-factor authentication (MFA), which has become nearly ubiquitous in high-security applications to combat the risk of credential stuffing.
As it stands, MFA increases the total entropy used to authenticate a user, but does so through mechanisms entirely independent of the primary authentication method. The goal of this paper is simply to incorporate the added entropy of MFA into password hashes to significantly increase their resistance to brute-force attacks without any added latency for users.

In this paper, we present techniques for building a multi-factor credential hashing function (MFCHF) that uses entropy from common authentication factors like HMAC-Based One-Time Password (HOTP) \cite{rfc4226}, Time-Based One-Time Password (TOTP) \cite{rfc6238}, and Out-Of-Band Authentication (OOBA) to strengthen password hashes without modifying existing client-side authentication hardware or software. Doing so in practice is non-trivial, as the dynamic nature of OTP factors is not readily conducive to their incorporation in a static hash. We overcome these limitations through a combination of novel cryptographic construction and techniques adapted from the field of multi-factor key derivation. The result is a dramatic improvement in brute-force attack resistance, with our experiments showing an MFCHF hash based on a password and HOTP would take over 8.5 years to crack compared to just 4.5 minutes with Argon2 alone (\S\ref{sec:bfr}). 

\subsubsection*{Contributions}

\begin{enumerate}[leftmargin=*]
    \item We performed an empirical study to demonstrate the real-world impact of hash function design on credential disclosure across over 4,000 prior data breaches (\S\ref{sec:stats}).
    \item We describe the first known method to use entropy from common authentication factors like HOTP, TOTP, OOBA, and YubiKeys within credential hashes (\S\ref{sec:mfchf}).
    \item Our scheme supports common usability features like validation windows (\S\ref{sec:validation}), factor persistence (\S\ref{sec:persistence}), and account recovery (\S\ref{sec:recovery}) with no loss in security.
    \item We demonstrate the merits of our approach through a full implementation (\S\ref{sec:pocapp}), and perform an experimental evaluation of its real brute-force attack resistance (\S\ref{sec:evaluation}).
\end{enumerate}

%% file: 020-Background.tex
\clearpage

\section{Background \& Motivation}

Although over \$150 billion is now spent annually on information security \cite{gartner_spending}, data breaches continue to experience exponential growth. In 2021 alone, there were over 4,000 publicly disclosed breaches containing over 22 billion records \cite{smag_breaches}. Thus, while the majority of current security research focuses on data breach prevention, the occurrence of thousands of data breaches per year should still be considered somewhat inevitable in today's security landscape. With the average cost of each data breach at an all-time high of \$4.35 million \cite{ibm_cost_2022}, minimizing risk in the event of a data breach is equally deserving of research attention. Indeed, the \textit{presume breach} tenet of the widely accepted zero-trust security framework compels implementers to give due consideration to harm reduction in the event that a data breach does occur \cite{zero_trust}.

While data breaches may contain a plethora of sensitive and personally identifiable information, password disclosure is amongst the most ubiquitous risks due to the prevalence of password-based authentication in user-facing systems.
Increasingly, the consequences of password disclosure extend not only to the directly affected systems, but also to unrelated third-party systems via credential stuffing due to widespread cross-site password reuse.
Hindering the ability of adversaries to obtain user credentials from breached databases thus remains a significant focus of post-breach risk mitigation efforts.

\subsection{Password Hashing}
\label{sec:hashing}

Password hashing is today considered the primary countermeasure to leaking passwords in the event of a data breach. Rather than storing passwords in plaintext, systems are configured to store a cryptographic hash of passwords corresponding to each user. Upon login, credentials presented by users are hashed and compared to the stored value. Since hash functions are non-reversible, the storage of password hashes avoids directly disclosing passwords in the event of a data breach.

Unfortunately, absent any further security considerations, hashed passwords are still highly susceptible to brute-force and dictionary attacks. While the average password contains just 40.54 bits of entropy \cite{florencio_large_2006}, modern hardware allows adversaries to calculate trillions of hashes per second for popular functions like SHA256 \cite{bitmain}, allowing password hashes to be reversed (``cracked'') and the plaintext password to be revealed, often in a matter of seconds.
Moreover, the deterministic nature of hash functions causes password reuse across users to be immediately evident even without cracking their hashes.

\subsubsection{Salting}
The practice of salting, whereby a randomly-generated ``salt'' value is used along with a password as input to a hash function and is then stored alongside the password, is considered best practice for all applications where password hashing is used. Its immediate consequence is adding a degree of non-determinism to the hashing process, such that even users with identical passwords will have different password hashes, thereby significantly slowing the process of cracking said hashes. In some instances, a further fixed random value, known as a ``pepper,'' is also used as an input to the hash function and is stored separately from the primary database.

\subsubsection{Brute-Force Resistance}
\begin{figure}[H]
\centering
\includegraphics[width=\linewidth]{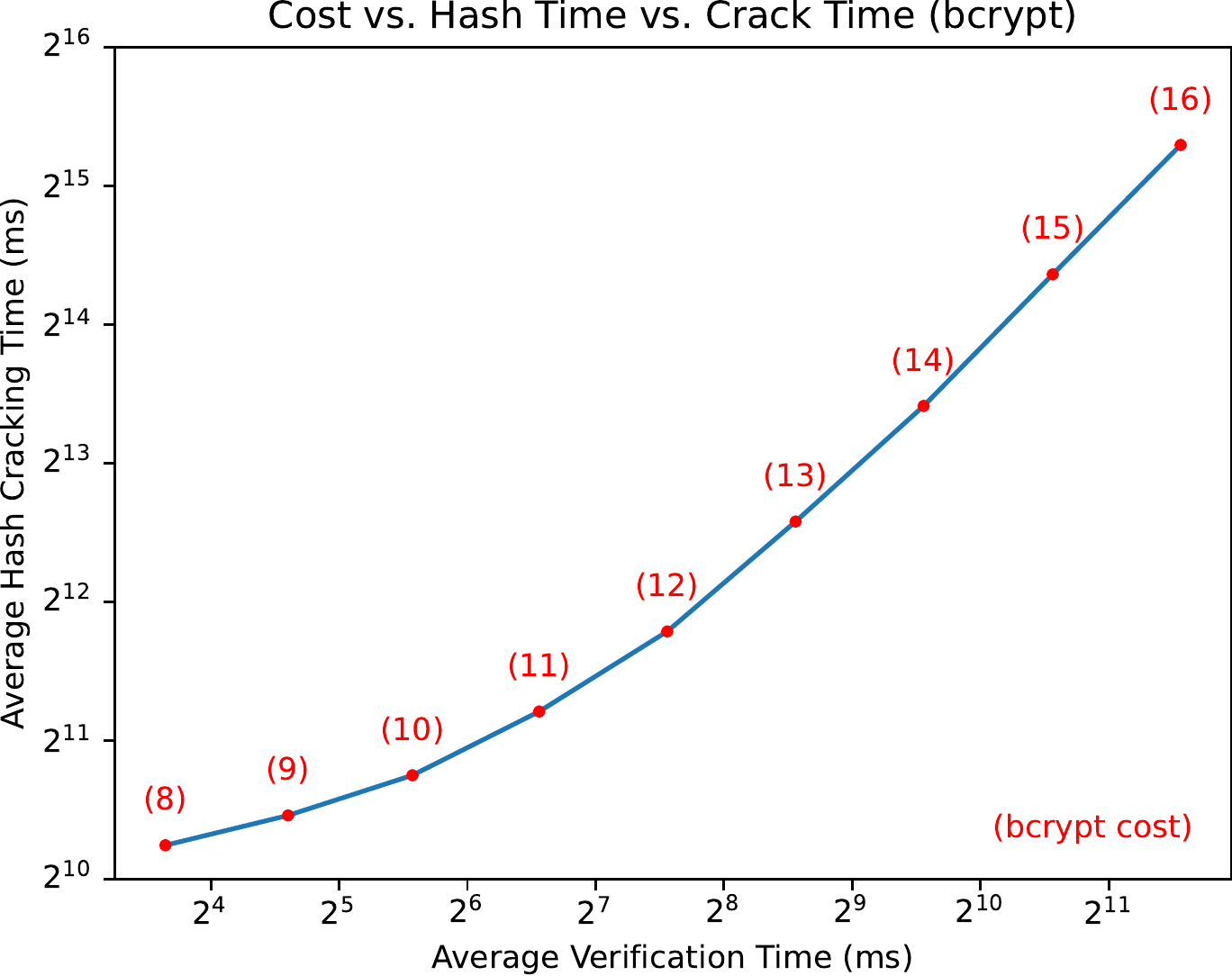}
\caption{\label{fig:bcrypt}Effect of bcrypt cost parameter on hash cracking time of an adversary and verification time for a user.}
\end{figure}

Even with the use of a salt, most cryptographic hash functions are optimized for computational efficiency, allowing brute-force attacks to proceed with relative speed. By contrast, most purpose-built password hashing functions are designed with a degree of intentional computational inefficiency, which significantly slows the pace of brute-force attacks at the cost of a slightly longer time to verify login attempts. In so-called ``adaptive'' password hashing functions, a variable cost parameter can be used to tune the function's computational difficulty and increase it over time as computing power improves. Fig.~\ref{fig:bcrypt} shows how changing the cost parameter of bcrypt \cite{bcrypt}, a popular adaptive password hashing function, affects the cracking time of an attacker and the verification time of a user.\footnote{Simulated cracking 10 salted bcrypt hashes corresponding to the top 100 passwords at each cost parameter. The hardware used to produce this graph and all other benchmarks in this paper is described in \S\ref{rig}.}

As illustrated in Fig.~\ref{fig:bcrypt}, a roughly linear relationship exists between verification time and hash cracking time for adaptive password hashing functions. In this paper, we categorize this relationship as \textit{symmetric resistance}, as increasing the difficulty of brute-force attacks necessarily accompanies a corresponding increase in the verification time for legitimate users. On the other hand, a technical improvement that increased the difficulty of cracking password hashes without impacting the verification time of a legitimate user would provide \textit{asymmetric resistance}.

A variety of password hashing mechanisms have been proposed with various approaches for providing symmetric resistance. In 2015, Argon2 \cite{argon2} was selected as the winner of a competition to determine the best hash function for hardware-resistant password hashing. 

\subsection{Breach Statistics}
\label{sec:stats}
While the choice of password hashing method theoretically plays an important role in preventing hash cracking after a data breach, how does the choice of hash function impact password disclosure in practice? We conducted a small measurement study to evaluate how password hashing has impacted the rate of password disclosure across thousands of real data breaches. To do so, we analyzed data from Hashmob, a ``password research and recovery'' community in which users around the world collaborate to crack hashes from over 4,000 verified data breaches comprising over 2.5 billion hashed credentials \cite{hashmob}.

\begin{figure}[H]
\centering
\includegraphics[width=\linewidth]{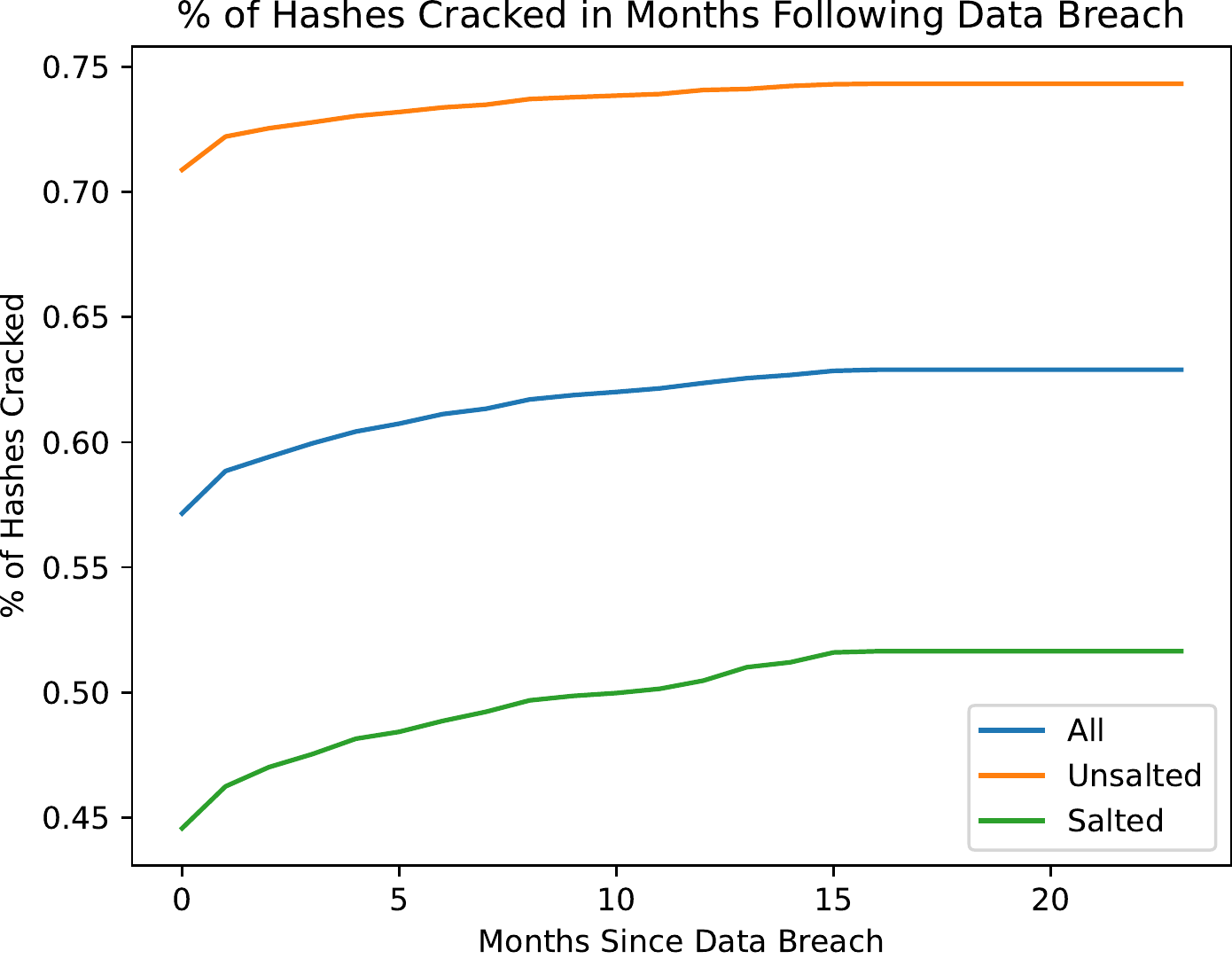}
\caption{\label{fig:curve}Progressive cracking of salted and unsalted hashes in the months following a data breach.}
\end{figure}

\subsubsection{Password Cracking Time}
Fig. \ref{fig:curve} illustrates the progression of a typical brute-force attack over time using Hashmob monthly progress data. In an average data breach, 57\% of password hashes are cracked within the first month, likely corresponding to those passwords which can easily be located via dictionary attacks. The rate of password cracking plateaus after the first year at around 74\% of hashes cracked, with no cracking attempts being reported more than 18 months after a data breach.

\begin{figure}[H]
\centering
\includegraphics[width=\linewidth]{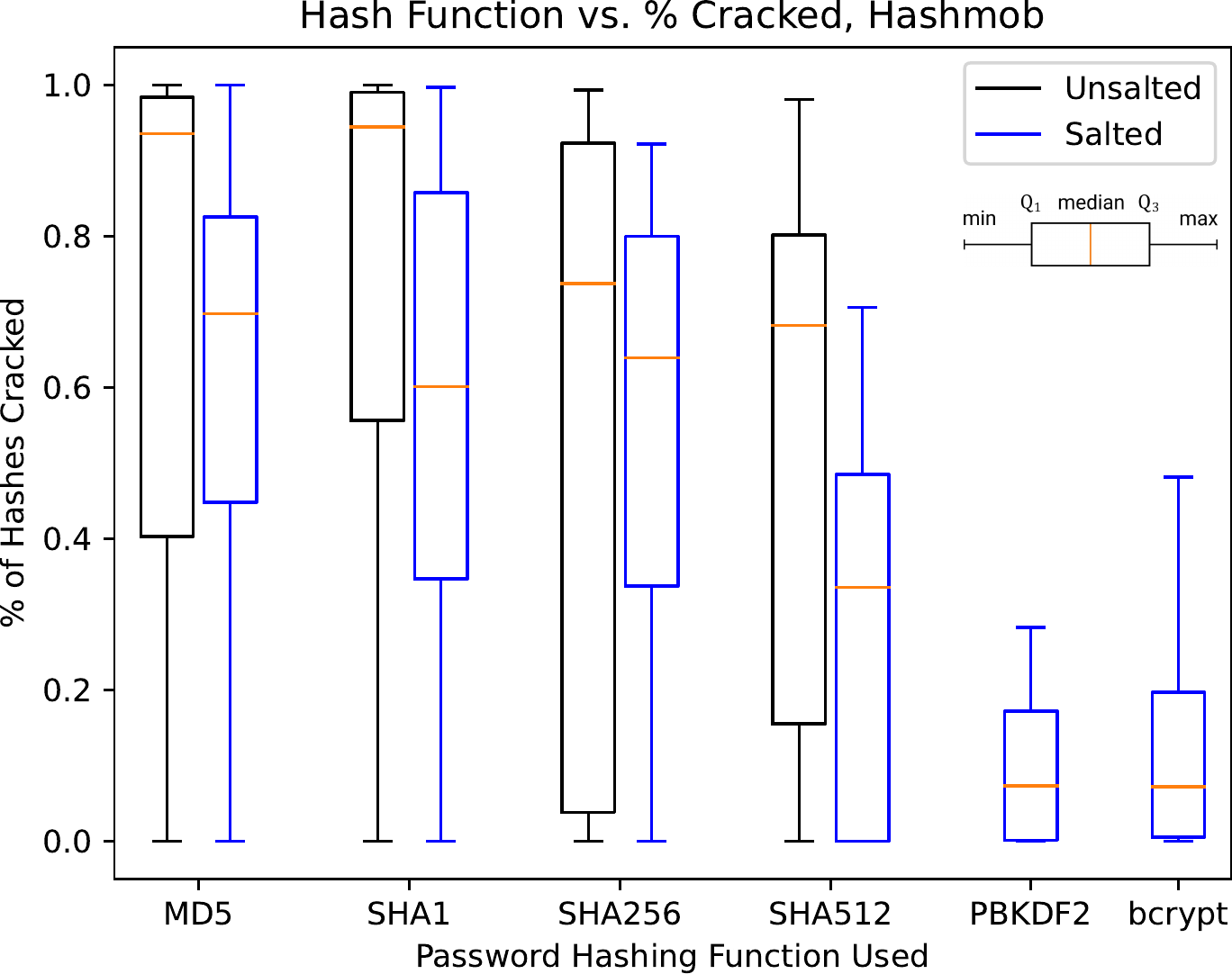}
\caption{\label{fig:ratio}Portion of hashes eventually cracked for hash types seen in data breaches on Hashmob (box plot).}
\end{figure}

\subsubsection{Password Disclosure Ratio}
\label{sec:ratio}
Having established that the vast majority of password-cracking activity occurs within the first year following a data breach, we analyzed all Hashmob data breaches posted at least one year ago to determine what effect the choice of hash function has on the percent of passwords eventually cracked by the community. The results for the six most popular password hashing methods are shown in Fig. \ref{fig:ratio}. 
The full results of our analysis of 4,259 data breaches are given in \S\ref{fulldata}.

\subsubsection{Key Findings}
We present the empirical findings of this section to emphasize that research into brute-force resistant password hashing methods has value that is not just theoretical but also clearly observable in the password cracking trends across thousands of actual data breaches. In particular, the beneficial effect of salting is clearly evident, with 51.6\% of salted hashes eventually being cracked, in comparison with 74.3\% of unsalted hashes. Even more clear is the impact of adaptive password hashing functions, with less than 15\% of bcrypt and PBKDF2 hashes being cracked even a full year after a data breach.

With an average per-record liability of \$164 \cite{ibm_cost_2022}, reducing the likely rate of password disclosure can have a significant impact on the level of risk associated with a data breach.
Overall, stronger hashing methods have the effect of increasing the average time between a data breach and credential disclosure, thereby giving companies more opportunity to detect and respond to the data breach, and of reducing the percentage of hashes eventually cracked, thereby potentially reducing the overall liability associated with the breach. Adaptive hash functions have clearly already had a massive positive impact in this regard.

Unfortunately, users are also highly sensitive to the latency of applications, with loading times of over 400~ms resulting in significantly decreased website traffic and conversion rate \cite{arapakis_impact_2021}.
Thus, there is effectively an upper limit on the extent to which symmetric brute-force resistance can be deployed without affecting usability. We thus argue that methods for achieving asymmetric resistance should be a significant focus of research in this area.

\subsection{Multi-Factor Authentication}
\label{sec:mfa}
The recent widespread adoption of multi-factor authentication (MFA) is owed in no small part to the incidence of password disclosure in data breaches, not because it currently serves to increase the difficulty of obtaining user passwords, but rather because it is largely implemented in response to the threat of credential stuffing attacks that result from leaked credentials.

While a variety of MFA mechanisms are currently in use, the most popular MFA methods in use today are one-time passwords (OTPs), such as HOTP and TOTP, and challenge-response mechanisms, such as HMAC-SHA1 (as used by YubiKeys). The effect of including these factors in the login process is basically to increase the overall entropy used to authenticate, as both the password and secondary factor are used together to verify the user. Below, we provide some background information on each of these schemes to clarify their integration with MFCHF.

\subsubsection{HMAC-SHA1}
HMAC-SHA1 challenge-response authentication is an instance of ISO/IEC 9798-2 2-Pass Unilateral Authentication via Cryptographic Check Function (CCF) \cite{ISO97982}, where the selected CCF is a Hash-based Message Authentication Code (HMAC) \cite{rfc2104} and Secure Hash Algorithm 1 (SHA1) \cite{rfc3174} is chosen as the underlying hash function. In a typical implementation, a client and server will share a 20-byte secret $\mathsf{key}$. The login process proceeds as shown below, where $\mathsf{HS1}(\mathsf{k},\mathsf{m})$ denotes HMAC-SHA1 with key $\mathsf{k}$ and message $\mathsf{m}$:

\begin{enumerate}[leftmargin=*]
    \item server $\rightarrow$ client: $\mathsf{challenge} \in [0,2^{160})$
    \item client $\rightarrow$ server: $\mathsf{response} = \mathsf{HS1}(\mathsf{key},\mathsf{challenge})$
    \item server: $\mathit{accept}$ iff $\mathsf{response} = \mathsf{HS1}(\mathsf{key},\mathsf{challenge})$
\end{enumerate}

\noindent Despite the deprecation of SHA-1 \cite{nist_deprecation} due to a lack of collision resistance, HMAC-SHA1 remains a secure \cite{bellare} and popular option for authentication due to its hardware-based support in products like YubiKey \cite{yubikey}, and forms the basis of most HOTP and TOTP implementations.

\subsubsection{HOTP}
HMAC-based one-time password (HOTP) \cite{rfc4226} is a 1-pass authentication mechanism that is typically based on HMAC-SHA1.\footnote{HOTP can be constructed using other underlying hash functions, but popular implementations like Google Authenticator only support SHA-1. Per the HOTP specification, a 31-bit truncation function is applied to the HMAC output before determining the final OTP value.} It replaces the challenge and response mechanism with a shared $\mathsf{counter}$ value that starts at $0$ and increments upon each successful login:

\begin{enumerate}[leftmargin=*]
    \item client $\rightarrow$ server: $\mathsf{OTP} = \mathsf{HS1}(\mathsf{key},\mathsf{counter})~\%~10^6$
    \item server: $\mathit{accept}$ iff $\mathsf{OTP} = \mathsf{HS1}(\mathsf{key},\mathsf{counter})~\%~10^6$
    \item client, server: increment $\mathsf{counter}$
\end{enumerate}

\noindent The elimination of a random challenge and the reduction of the response size to a small number of decimal digits (usually 6) has made HOTP a popular choice for smartphone-based 2FA apps like Google Authenticator.

\subsubsection{TOTP}
Time-based one-time password (TOTP) \cite{rfc6238} is an extension of HOTP that replaces the shared counter with a coarse timestamp. As with HOTP, it is typically based on HMAC-SHA1. Given the current UNIX time $T$, an initial time $T_0$, and a time interval $T_X$, the TOTP code at time $T$ is equal to $\mathsf{HOTP_K}(\lfloor(T-T_0)/T_X\rfloor)$. TOTP has the advantage of avoiding the counter desynchronization issues of HOTP.\\

HOTP, TOTP, and HMAC-SHA1 all essentially serve to supplement passwords with additional entropy derived from a key using HMAC, increasing the overall entropy used to authenticate a user.
The goal of this paper is simply to take advantage of this added entropy in the password hashing process to increase the brute-force difficulty of the resulting hash while, assuming MFA was already in use, having no significant impact on the user experience, thus providing asymmetric brute-force resistance.
Achieving this would require popular OTP authentication methods to be incorporated into the hashing mechanism without modifying the client-side functionality of these factors, such that the user experience remains largely unaffected.

Unfortunately, two major difficulties remain in the realization of this technique. Firstly, while passwords, and thus password hashes, remain fairly constant over time, OTPs are by definition intended for one-time use, and are thus expected to change upon each login. It may not be immediately clear how a static hash can be guaranteed to reflect the OTP corresponding to any given login request.

Secondly, while the server must retain the ability to validate all authentication factors, it can no longer centrally store secret information about those factors. For example, HOTP and TOTP codes are typically verified by storing a shared HMAC secret key in the database along with a password hash, but doing so would defeat the purpose of using said OTP as part of the hash, as an adversary obtaining a copy of the database could easily reproduce the correct OTP and defeat any added difficulty.

Thus, while taking advantage of multi-factor authentication to increase the difficulty of cracking password hashes seems straightforward, doing so in practice is easier said than done and requires the design of new techniques.

\subsection{MFKDF}
The Multi-Factor Key Derivation Function (MFKDF) \cite{mfkdf} is a recent improvement over password-based key derivation that incorporates multiple authentication factors into the key derivation process. Its construction provides an important building block for the creation of a multi-factor credential hashing function with support for commonly-used OTP authentication factors.

The MFKDF specification contains two major architectural components. The first component is the set of so-called ``factor constructions,'' which convert a dynamic \textit{factor witness}\footnote{The \textit{witness} refers to the message used to authenticate (e.g., a 6-digit OTP), which is often not the same as the underlying shared secret.} ($W$) and public parameters ($\alpha$) into static key material ($\sigma$). The public parameters require no security assumptions and can safely be stored in a database without concern for revealing information about the factors to potential adversaries. For some factors, these parameters must be updated upon each login ($\alpha_i \mapsto \alpha_{i+1}$). Constructions are given for a variety of popular authentication factors, including TOTP, HOTP, OOBA (e.g., Email/SMS), and HMAC-SHA1 (e.g., YubiKey).

The second major component of the MFKDF specification is the key derivation function itself, which adds a secret sharing layer to provide functionality such as threshold-based key derivation, advanced policy enforcement, and factor recovery. While useful in the key derivation setting, such functionality is not particularly relevant for building a multi-factor credential hashing function and adds unnecessary overhead in the form of needing to store excessive material like secret shares.

While the intended use case of MFKDF, namely client-side key derivation for end-to-end encryption, is very different from the goals of this paper, the MFKDF factor constructions are a key tool for solving the challenge of using dynamic OTP factors as an input to a static hash. The core technique of this paper builds atop said factor constructions to achieve multi-factor credential hashing with asymmetric resistance.

\subsection{Summary}

With the introduction of ultra-fast hashing ASICs for cryptocurrency mining, research into brute-force-resistant password hashing mechanisms has become more important than ever before. We hope the brief empirical study presented thus far serves to demonstrate that advancements in this area have a dramatic effect on the consequences of actual data breaches.

Clearly, the symmetric resistance provided by adaptive password hashing functions has already had a significant impact on password cracking, but the relative intolerance of users to added latency places an upper limit on the potential use of these functions. Simultaneously, the adoption of multi-factor authentication has provided a key opportunity to incorporate additional entropy into credential hashes without increasing the computation time for legitimate users.
We are thus motivated to explore a scheme that takes advantage of the additional entropy provided by MFA to provide asymmetric resistance to brute-force attacks by incorporating multiple authentication factors, rather than just passwords, into a single hash.

While achieving such a scheme has long seemed out of reach, the introduction of MFKDF has provided a blueprint for realizing its implementation.
In the following section, we will outline the desired security properties of our multi-factor credential hashing function, and will then proceed to describe a scheme satisfying these properties.

%% file: 030-Problem-Statement.tex
\section{Problem Statement}

In a typical password hashing deployment, a user registers an account with a server by providing a password, a hash of which is stored by the server in a database. Later, the user initiates a login process with a server by providing the password, which the server hashes and compares to the stored hash to authenticate the request. Secondary authentication factors are then independently verified.

In this paper, we'll present a multi-factor credential hashing approach that differs from typical password hashing by atomically verifying all of a user's credentials with a single hash. The purpose of this section is to present the problem setting and goals of this approach. These are largely the same as in standard password hashing, except that all factors are provided and verified simultaneously rather than sequentially, yielding a  significant asymmetric improvement in brute-force attack resistance.

\subsection{Deployment Setting}
\noindent Our deployment setting consists of the following entities:

\begin{itemize}[leftmargin=*]
    \itemsep 0em
    \item One or more \textit{users}, each possessing a password along with a secondary authentication factor.
    \item A \textit{server}, which stores data about the factors of each user and uses it to authenticate user login requests.
\end{itemize}

\subsection{Registration Process}
In an initial setup process, a user establishes a password with the server. The server establishes a secondary authentication factor and internally stores a hash relating to these factors that can be used to later authenticate the user. Thus, the MFCHF $\textsc{Setup}$ function requires the following type definition: \\

${\textsc{Setup}}: {\mathsf{password}\mapsto\mathsf{hash},\mathsf{mfainfo}}$

\subsection{Login Process}
During a login process, a user simultaneously provides the server with witnesses corresponding to all authentication factors. For example, they may provide a password along with a TOTP code (e.g., from Google Authenticator). The server must then use the data stored during the registration process to validate all factors and authenticate the user. If authentication succeeds, the server may also update its stored hash to prepare for the next login. Thus, the type definition for the MFCHF $\textsc{Verify}$ function is as follows: \\

${\textsc{Verify}}: {\mathsf{password},\mathsf{witness},\mathsf{hash}\mapsto\mathit{reject}}$ or \\
\indent\indent\indent\indent\indent\indent\indent\indent\indent~~~~~ ${\mapsto\mathit{accept},\mathsf{hash}}$

\medskip

\subsection{Threat Model}
\label{sec:threat-model}
We consider security under a \textit{total data breach} threat model; that is, at some instant, an adversary receives a snapshot of all data stored on the server. The goal of the adversary is to use this data to obtain the underlying credentials of the user, via brute force, dictionary attack, or any other method available to the adversary in an offline capacity. To ensure a fair comparison with other password hashing schemes, the adversary is considered successful even if just the user's password is determined.

If given unlimited time, the adversary will be able to brute-force user credentials under any scheme where the stored data allows credentials to be verified. Therefore, the goal of our scheme is to increase the time and difficulty of attacking user credentials without increasing the time taken to verify a legitimate user (asymmetric resistance).

\medskip

\subsection{Security Goals}
\label{sec:goals}
\noindent To summarize, a multi-factor credential hashing function consists of separate $\textsc{Setup}$ and $\textsc{Verify}$ functions. The security goals of an MFCHF scheme are as follows: \\

\begin{enumerate}[leftmargin=*]
    \itemsep 1em
    \item \textbf{Correctness:} When provided with valid witnesses corresponding to a user's established authentication factors and hash, the server outputs $\mathit{accept}$ with $p=1$.
    \item \textbf{Safety:} When at least one of a user's factor witnesses is invalid with respect to the established factors and hash, the server outputs $\mathit{reject}$ except with $p=negl$.
    \item \textbf{Asymmetric Resistance:} For a given fixed verification time, a multi-factor hash should be significantly more difficult to brute-force than a standard password hash.
\end{enumerate}

%% file: 040-MFCHF.tex
\section{Multi-Factor Credential Hashing}
\label{sec:mfchf}
\vspace{-0.7em}
We now present four practical constructions for two-factor credential hashing schemes corresponding to authentication using a password and either HOTP, TOTP, OOBA, or HMAC-SHA1 (YubiKey) as a secondary factor. While these schemes can easily be extended to arbitrary $n$-factor variants, their construction is most straightforwardly presented using two factors, which is the most common use case.
Furthermore, our four chosen secondary factors are by no means exhaustive with respect to the authentication methods that can be used with multi-factor credential hashing, and exist to serve as a template for implementing the proposed approach with other authentication methods one may wish to use in the future while simultaneously demonstrating the backward compatibility of MFCHF with several popular unmodified authentication factors.

\subsection{General Blueprint}
\label{sec:blueprint}
\begin{figure}[H]
\centering
\includegraphics[width=\linewidth]{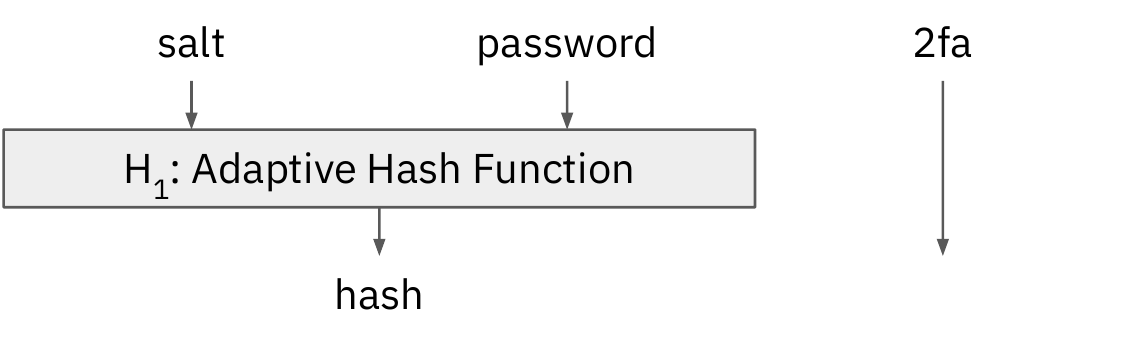}
\caption{\label{fig:blueprint1}Standard adaptive password hashing scheme.}
\end{figure}

We begin with an overview of our general approach for multi-factor credential hashing. Fig. \ref{fig:blueprint1} shows a typical password hashing setup using a salted adaptive hash function. Even if MFA is in use, secondary factors are considered totally independent of password validation.

\vspace{-0.5em}

\vspace{-0.5em}

As discussed in \S\ref{sec:mfa}, our goal is to incorporate the additional entropy of a secondary authentication factor into the hash function as though it were an additional hidden salt value, but we are unable to do so directly due to the dynamic nature of OTPs. A successful multi-factor credential hashing scheme must convert a dynamic secondary factor into another static input, and must further obscure the private material underlying that factor.

\vspace{-0.5em}

\begin{figure}[H]
\centering
\includegraphics[width=\linewidth]{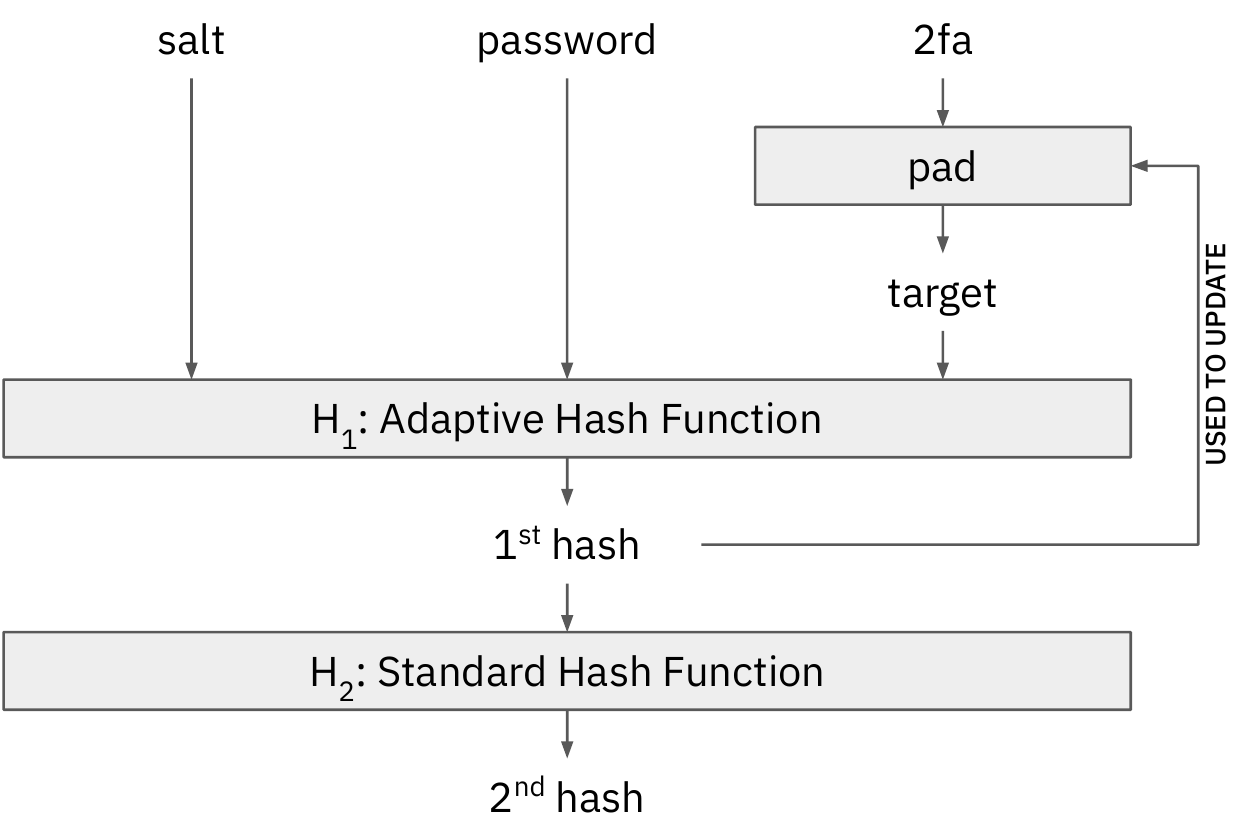}
\caption{\label{fig:blueprint2}Blueprint for multi-factor credential hashing.}
\end{figure}

\vspace{-0.5em}

Fig. \ref{fig:blueprint2} illustrates the general strategy implemented by all of our MFCHF constructions. It incorporates three major insights: First, a one-time pad or modular offset can be used to convert dynamic OTPs into a static hash input without leaking information. Second, an intermediate hash value can be used as a key to hide secrets within the data stored on the server. Finally, those secrets can be recovered ephemerally upon each login to update the pad or offset.

The resulting feedback loop incorporates dynamic secondary factors into the hash without weakening them, and actually strengthens the underlying construction by not storing shared keys in plaintext.
The forthcoming MFCHF constructions all start with this general blueprint and then introduce their own factor-specific optimizations.

\subsection{MFCHF with HOTP (mfchf-hotp6)}
\label{sec:hotp}

\vspace{-0.7em}

Our first specific MFCHF construction uses HOTP, a factor that is illustrative of the general approach described in \S\ref{sec:blueprint}.
The mfchf-hotp6 construction requires two underlying hash functions: $\mathsf{H_1}$ is an adaptive password hash function such as Argon2, and $\mathsf{H_2}$ is a standard cryptographic hash function such as SHA-256. We assume that a $\mathsf{salt}$ and HOTP $\mathsf{key}$ have been randomly chosen ahead of time.
$\mathsf{HOTP}(\mathsf{k},\mathsf{n})$ denotes the n\textsuperscript{th} HOTP code under key $\mathsf{k}$ per RFC 4226 \cite{rfc4226}.
Throughout this paper, $\oplus$ denotes bitwise XOR and $\odot$ denotes concatenation. The \textsc{Setup} function for mfchf-hotp6 then proceeds as follows:

\begin{enumerate}[leftmargin=*]
    \item Select a random $\mathsf{target}$ value in $[0,10^6)$
    \item Determine the first HOTP code: $\mathsf{first}=\mathsf{HOTP}(\mathsf{key}, 1)$
    \item Find the modular offset: $\mathsf{diff} = (\mathsf{target} - \mathsf{first})~\%~10^{6}$
    \item Compute 1\textsuperscript{st} hash: $\mathsf{inner} = \mathsf{H_1}(\mathsf{password} \odot \mathsf{target} \odot \mathsf{salt})$
    \item Blind HOTP key using 1\textsuperscript{st} hash: $\mathsf{blind} = \mathsf{key} \oplus \mathsf{inner}$
    \item Compute 2\textsuperscript{nd} hash: $\mathsf{outer} = \mathsf{H_2}(\mathsf{inner})$
    \item Store $\{\mathsf{counter}=1,\mathsf{diff},\mathsf{blind},\mathsf{salt},\mathsf{outer}\}$
\end{enumerate}

\noindent Upon login, the user obtains an $\mathsf{otp}$ code from their HOTP application, which is provided to the server and used along with their password within the \textsc{Verify} function like so:

\begin{enumerate}[leftmargin=*]
    \item Recover $\mathsf{target}$ value: $\mathsf{target} = (\mathsf{diff} + \mathsf{otp})~\%~10^6$
    \item Compute 1\textsuperscript{st} hash: $\mathsf{inner} = \mathsf{H_1}(\mathsf{password} \odot \mathsf{target} \odot \mathsf{salt})$
    \item Output $\mathit{reject}$ if $\mathsf{H_2}(\mathsf{inner}) \neq \mathsf{outer}$, else continue
    \item Increment $\mathsf{counter}$ value 
    \item Unblind HOTP key using 1\textsuperscript{st} hash: $\mathsf{key} = \mathsf{blind} \oplus \mathsf{inner}$
    \item Determine the next OTP: $\mathsf{next}=\mathsf{HOTP}(\mathsf{key}, \mathsf{counter})$
    \item Find the modular offset: $\mathsf{diff} = (\mathsf{target} - \mathsf{next})~\%~10^{6}$
    \item Output $\mathit{accept}$, store $\{\mathsf{counter},\mathsf{diff},\mathsf{blind},\mathsf{salt},\mathsf{outer}\}$
\end{enumerate}

\noindent Because the original $\mathsf{target}$ value is recovered only if the $\mathsf{otp}$ value is correct, the correct password and OTP must be provided in order for the server to output $\mathit{accept}$. Thus, the MFCHF hash has the effect of simultaneously validating the password and HOTP factors. See Alg. \ref{alg:mfchf_hotp6} of \S\ref{app:algs} for a pseudocode implementation of this method.

\subsubsection{HOTP Validation Window}
\label{sec:validation}

A common usability feature in HOTP-based systems is the use of a ``validation window'' to recover from counter desynchronization.
This functionality can easily be added onto the above HOTP construction by storing multiple offset ($\mathsf{diff}$) values corresponding to multiple acceptable counter values, and iteratively trying each of the stored offsets upon login. For example, if a validation window of size 2 is desired, the offsets corresponding to $\mathsf{ctr}$ and $\mathsf{ctr}+1$ are stored at all times.
Because each offset must be used independently, including additional offsets does not reduce the security of the scheme, and instead simply increases the verification time if multiple attempts are required. For instance, even if one stored $10^6$ offsets, trying each of them would be no faster than trying all $10^6$ possible target values.

\subsection{MFCHF with TOTP (mfchf-totp6)}
\label{sec:totp}

\vspace{-0.5em}

Per RFC 6239, ${\mathsf{TOTP_K}=\mathsf{HOTP_K}(\lfloor(T-T_0)/T_X\rfloor)}$ where $T$ is the current UNIX time, $T_0$ is the initial time, and $T_X$ is the time interval  \cite{rfc6238}. Accordingly, the above MFCHF construction for HOTP can be modified to produce a suitable construction for TOTP by storing an array of offset ($\mathsf{diff}$) values corresponding to the next $w$ OTPs. Because the underlying HMAC key is available to the server in plaintext within the \textsc{Setup} and \textsc{Verify} functions, this calculation can still be performed locally with no required modifications to the TOTP application.
A full description of mfchf-totp6 is given in Alg. \ref{alg:mfchf_totp6} of \S\ref{app:algs}.

\subsection{MFCHF with OOBA (mfchf-ooba6)}
\label{sec:ooba}

\vspace{-0.5em}

We next turn our attention to out-of-band authentication (OOBA) factors such as email and SMS. We assume the use of public key cryptography underlying each OOBA channel, which can be used to deliver an OTP only to an intended recipient.
Let $\mathsf{Enc}$ represent encryption under a public-key encryption scheme (e.g., RSA) and $\mathsf{pk}$ represent the public key of the OOBA channel. As before, we assume that a $\mathsf{salt}$ has been randomly chosen ahead of time, but only a single password hash function ($\mathsf{H}$) is required. The \textsc{Setup} function is thus as follows:

\begin{enumerate}[leftmargin=*]
    \item Select a random $\mathsf{first}$ OTP and $\mathsf{target}$ value in $[0,10^6)$
    \item Encrypt the $\mathsf{first}$ OTP with $\mathsf{pk}$: $\mathsf{ct}=\mathsf{Enc}(\mathsf{first},\mathsf{pk})$
    \item Find the modular offset: $\mathsf{diff} = (\mathsf{target} - \mathsf{first})~\%~10^{6}$
    \item Compute the hash: $\mathsf{hash} = \mathsf{H}(\mathsf{password} \odot \mathsf{target} \odot \mathsf{salt})$
    \item Store $\{\mathsf{ct},\mathsf{pk},\mathsf{diff},\mathsf{salt},\mathsf{hash}\}$
\end{enumerate}

\noindent Because our OOBA implementation is not limited by backward compatibility with existing HOTP/TOTP software, it is not restricted to using numeric OTPs. Instead, base 36 (or even 62) representation can be used to provide alphanumeric OTPs. Upon login, the server can forward $\mathsf{ct}$ to the OOBA channel. The user then provides the received $\mathsf{otp}$ to the server along with their $\mathsf{password}$, which are used within the \textsc{Verify} function like so:

\begin{enumerate}[leftmargin=*]
    \item Recover $\mathsf{target}$ value: $\mathsf{target} = (\mathsf{diff} + \mathsf{otp})~\%~10^6$
    \item Output $\mathit{reject}$ if $\mathsf{H}(\mathsf{password} \odot \mathsf{target} \odot \mathsf{salt}) \neq \mathsf{hash}$
    \item Select a random $\mathsf{next}$ OTP value in $[0,10^6)$
    \item Encrypt the $\mathsf{next}$ OTP with $\mathsf{pk}$: $\mathsf{ct}=\mathsf{Enc}(\mathsf{next},\mathsf{pk})$
    \item Find the modular offset: $\mathsf{diff} = (\mathsf{target} - \mathsf{next})~\%~10^{6}$
    \item Output $\mathit{accept}$, store $\{\mathsf{ct},\mathsf{pk},\mathsf{diff},\mathsf{salt},\mathsf{hash}\}$
\end{enumerate}

\noindent Again, the server accepts the authentication request only if the user provides the correct $\mathsf{otp}$, indicating that they had access to the OOBA channel. As in MFKDF, the recommended implementation of the OOBA factor for email authentication is to use the S/MIME key \cite{rfc3850} of the recipient as $\mathsf{pk}$. This can be extended to SMS authentication using the \mbox{email-to-SMS} gateway service \cite{sms_gateway} of each carrier. Alg. \ref{alg:mfchf_ooba6} of \S\ref{app:algs} further describes mfchf-hotp6.

\subsection{MFCHF with YubiKey (mfchf-hsha1)}
\label{sec:hsha1}

\vspace{-0.5em}

Finally, we provide an MFCHF construction for hardware-based MFA devices such as smart cards and USB security keys. Amongst the most common protocols supported by these devices are FIDO U2F \cite{u2f} and ISO 9798 2-Pass Unilateral Authentication over HMAC-SHA1 \cite{ISO97982}. Unfortunately, the former is effectively impossible to incorporate into a multi-factor credential hash due to the inclusion of a client-side random nonce in all signatures. However, HMAC-SHA1 is both well supported (including all YubiKey devices \cite{yubikey}) and relatively straightforward to implement as an MFCHF factor. Our method requires a single password hash function ($\mathsf{H}$) and HMAC-SHA1 ($\mathsf{HS1}$). Assuming a $\mathsf{key}$ and $\mathsf{salt}$ have already been chosen, the \textsc{Setup} function for mfchf-hsha1 is as follows:

\begin{enumerate}[leftmargin=*]
    \item Select a random $\mathsf{challenge}$ in $[0,2^{160})$
    \item Find the $\mathsf{response}$: $\mathsf{response}=\mathsf{HS1}(\mathsf{key},\mathsf{challenge})$
    \item Compute the hash: $\mathsf{hash} = \mathsf{H}(\mathsf{password} \odot \mathsf{key} \odot \mathsf{salt})$
    \item Blind key using response: $\mathsf{blind} = \mathsf{key} \oplus \mathsf{response}$
    \item Store $\{\mathsf{blind},\mathsf{challenge},\mathsf{salt},\mathsf{hash}\}$
\end{enumerate}

\noindent Upon login, the $\mathsf{challenge}$ can be sent to the user's hardware device to generate a $\mathsf{response}$, which is used along with their $\mathsf{password}$ within the \textsc{Verify} function like so:

\begin{enumerate}[leftmargin=*]
    \item Unblind key using response: $\mathsf{key} = \mathsf{blind} \oplus \mathsf{response}$
    \item Output $\mathit{reject}$ if $\mathsf{H}(\mathsf{password} \odot \mathsf{key} \odot \mathsf{salt}) \neq \mathsf{hash}$
    \item Select a random $\mathsf{challenge}$ in $[0,2^{160})$
    \item Find the $\mathsf{response}$: $\mathsf{response}=\mathsf{HS1}(\mathsf{key},\mathsf{challenge})$
    \item Blind key using response: $\mathsf{blind} = \mathsf{key} \oplus \mathsf{response}$
    \item Output $\mathit{accept}$, store $\{\mathsf{blind},\mathsf{challenge},\mathsf{salt},\mathsf{hash}\}$
\end{enumerate}

\noindent The authentication request is accepted only if the user is in possession of the hardware MFA device containing the shared $\mathsf{key}$ and is thus able to produce the correct $\mathsf{response}$. The full specification is given in Alg. \ref{alg:mfchf_hmacsha1} of \S\ref{app:algs}.

\subsection{Use of SHA-1}
\label{sec:sha1}

\vspace{-0.5em}

Because SHA-1 has been deprecated since 2011 due to its lack of collision resistance \cite{nist_deprecation}, it may be seen as a red flag to recommend its use in new cryptographic deployments.
However, the security of HMAC is not dependent upon collision resistance, and HMAC-SHA1 has been proven to remain secure without it \cite{bellare}. 

Moreover, our hand is forced by the exclusive use of SHA-1 in popular MFA methods. For instance, HMAC-SHA1 remains the only deterministic challenge-response mechanism supported by YubiKeys. Furthermore, while HOTP and TOTP support the use of other hash functions, Google Authenticator still only supports HMAC-SHA1.

\subsection{Post-Breach Security}
\label{sec:postbreach}

\vspace{-0.5em}

Each of the four MFCHF constructions introduced in this section has the primary goal of increasing the brute-force resistance of stored password hashes by leveraging the entropy of MFA within stored hashes. For example, the HOTP, TOTP, and OOBA variants require the attackers to search the entire space of $\{\mathsf{password} \odot \mathsf{target}\}$ instead of just $\{\mathsf{password}\}$. This aspect of MFCHF's security is evaluated in detail in \S\ref{sec:evaluation}. However, the MFCHF techniques presented herein have the additional advantage of not storing factor-specific secrets in plaintext. As a result, the underlying authentication factors have the potential to remain operational in the event of a data breach. For example, a typical HOTP/TOTP-based system would store the HMAC key in plaintext on the server to verify OTPs, thus leaking the key in the event of a data breach and allowing the attacker to bypass the HOTP/TOTP factor entirely. On the other hand, MFCHF allows the server to verify OTPs without storing the HMAC key in plaintext, thus potentially leaving the factor operational in the period between the occurrence of a data breach and its detection.

%% file: 050-Features.tex
\clearpage

\section{Authentication Features}
\label{sec:features}

A major focus of our discussion thus far has been on the backward compatibility of MFCHF with existing authentication hardware and software, so as to emphasize that MFCHF requires modifications to neither third-party applications nor learned user behaviors. Similarly, in \S\ref{sec:validation}, we demonstrated that HOTP validation windows, a commonly-implemented feature designed to improve usability, are fully compatible with the mfchf-hotp6 construction.
In general, a significant goal of this paper is to reduce the friction of implementing MFCHF to minimize the drawbacks accompanying its significant security \mbox{advantages} and thereby satisfy the balance of considerations preceding its implementation.

To that end, there are two additional usability features that an authentication scheme must support to avoid a heavy usability penalty: factor persistence and factor recovery. The aim of this section is to illustrate that these features can easily be implemented in conjunction with the proposed MFCHF methods while not reducing the security or brute-force resistance of the underlying schemes. In the following section, these features are implemented together with the mfchf-hotp6 scheme to produce a proof-of-concept application that realizes the security benefits of MFCHF while retaining all popular usability features.

\subsection{Factor Persistence}
\label{sec:persistence}

Factor persistence is a common usability feature that allows users to bypass multi-factor authentication when using a familiar trusted device. A standard method of implementing factor persistence is by storing a browser cookie containing a pseudorandom token value on devices the user indicates are trusted. Login requests containing the cookie with the correct value are only required to supply the primary authentication factor, bypassing any multi-factor authentication that may be in use. Because each factor is typically validated by an entirely independent mechanism, implementing factor persistence usually requires fairly limited additional server-side logic. However, when using MFCHF, implementing factor persistence in the usual manner would require separate storage of a password-only hash to facilitate primary authentication, thereby defeating the brute-force resistance of MFCHF.

Fortunately, each of the presented MFCHF constructions contains a built-in token that can be used to achieve factor persistence, namely the $\mathsf{hmackey}$ in the case of mfchf-hsha1, and the $\mathsf{target}$ value in the case of HOTP, TOTP, and OOBA. While these values are never stored in plaintext, they become temporarily available within each of the \textsc{Verify} functions upon login. Thus, when a user successfully authenticates using a new device, they can be prompted to bypass MFA on that device, storing their $\mathsf{hmackey}$ or $\mathsf{target}$ value as a cookie in the process. On subsequent logins, that value can be used together with their password to authenticate their request without necessitating the use of multi-factor authentication. Because only a multi-factor hash is stored, this solution allows the use of factor persistence with MFCHF with no loss in security or brute-force resistance as long as the trusted device is secure (which is the assumption fundamentally made by allowing a given device to bypass MFA).

\subsection{Factor Recovery}
\label{sec:recovery}

Factor recovery is another usability feature that is vital to support, with nearly 80\% of users requiring a password reset on a regular basis \cite{password_reset}. Typically, factor recovery is implemented by establishing a tertiary factor, such as a recovery code or email OOBA, specifically for the purpose of recovering a lost primary or secondary factor. For example, in a password plus HOTP scheme, the password and recovery code may be used together to recover a lost HOTP device. Once again, a naive implementation of this setup may involve storing a separate password hash, thereby defeating the security of MFCHF.

Here, we suggest an alternative solution that involves storing a separate MFCHF hash for each combination of factors an application wishes to support. We will specifically follow the example of password and HOTP authentication with a recovery code, but stress that any supported factor can be used for recovery via this method. Implementing HOTP device recovery in this scheme is easily achieved by storing a hash of the password and recovery code. Password recovery, however, is non-trivial to implement, and can be achieved as follows:

\subsubsection*{Setup}
During the setup process, a password recovery hash is created by applying a password hash function to a recovery code concatenated with the $\mathsf{target}$ value of the primary mfchf-hotp6 authentication hash. The $\mathsf{hotpsecret}$ should be stored padded with the output of this recovery hash in addition to the primary hash. As with the primary hash, a standard cryptographic hash function like SHA256 should then be applied before storing the resulting value.

\subsubsection*{Verify}
The $\mathsf{counter}$ and $\mathsf{offset}$ values from the \mbox{primary} mfchf-hotp6 hash should be synchronized with the \mbox{recovery} hashes upon each successful login to ensure the same HOTP factor used for primary authentication can be used to recover the account password if required.

\subsubsection*{Recovery}
During an account recovery process, the HOTP code can be combined with the $\mathsf{offset}$ value to recover the $\mathsf{target}$ as usual, which in turn can be verified together with the supplied recovery code against the recovery hash.
Because the $\mathsf{hotpsecret}$ is stored padded with the recovery hash, it can be found when the recovery hash is computed.

\subsubsection*{Reconstitution} Finally, the known $\mathsf{hotpsecret}$ and $\mathsf{target}$ values can be combined with a new password to create a new mfchf-hotp6 hash for primary authentication while leaving the HOTP factor and recovery hash intact. \\

The proposed method for factor recovery averts storing an independent hash for each authentication factor and instead only stores multi-factor hashes for allowable factor combinations. As such, it implements factor recovery via tertiary factors without reducing the security of the overall scheme. Specifically, it reduces the overall brute-force difficulty only to the weakest combination of factors allowed to authenticate, which is the strongest possible security a multi-factor credential hashing scheme can hope to achieve. This principle is further discussed in \S\ref{sec:pocentropy}, which evaluates the overall brute-force entropy of an MFCHF demo application with three MFCHF hashes.

%% file: 090-Demo.tex
\clearpage

\section{Proof-of-Concept Implementation}
\label{sec:pocapp}

To demonstrate the immediate practical utility of MFCHF and provide a blueprint for its deployment, we implemented a fully-featured MFCHF JavaScript library, and produced a proof-of-concept application implementing MFCHF as a replacement for password hashing. The demo application is modeled as a template, built using a React.js frontend and serverless JavaScript backend, with full authentication functionality (registration, login, recovery, etc.) but no substantive application content.

\subsection{MFCHF Deployment}

In our proof-of-concept application, passwords are used as a primary authentication factor with HOTP as a secondary factor. Therefore, the primary hash used for authentication reflects the mfchf-hotp6 algorithm described in \S\ref{sec:hotp}, with PBKDF2 used as the underlying password hash in this instance. We verified that the resulting application was fully backward-compatible with the latest versions of the Google Authenticator and Microsoft Authenticator applications from the Google Play store. As such, the use of MFCHF for hashing on the backend should have no discernible impact on users who already use HOTP, other than requiring all factors to be entered simultaneously rather than sequentially as seen in Fig. \ref{fig:login}.

\subsection{Authentication Features}

\begin{figure}[H]
    \centering
    \begin{subfigure}[b]{0.49\linewidth}
        \centering
        \includegraphics[width=\textwidth]{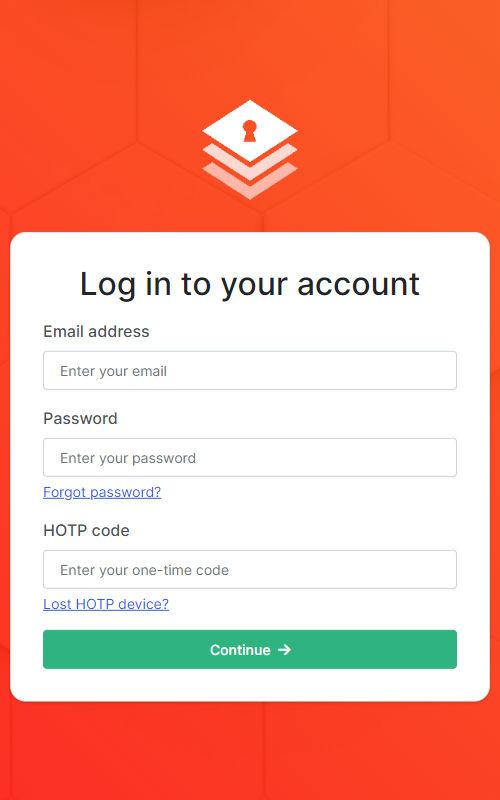}
        \caption{Login}
        \label{fig:login}
    \end{subfigure}
    \hfill
    \begin{subfigure}[b]{0.49\linewidth}
        \centering
        \includegraphics[width=\textwidth]{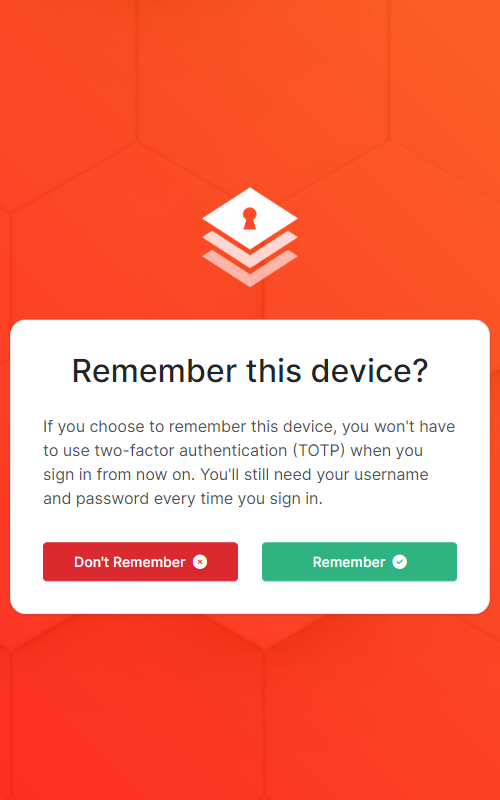}
        \caption{Persistence}
        \label{fig:persistence}
    \end{subfigure}
    \caption{Login screens from proof of concept application.}
    \label{fig:demo}
\end{figure}

To emphasize the limited usability impact of MFCHF, we implemented a number of common usability and convenience features within the proof of concept app, demonstrating their practical compatibility with MFCHF.
In particular, the application supports the use of a HOTP validation window (i.e., a look-ahead window) using the method of \S\ref{sec:validation}, allowing users to recover from the desynchronization of their HOTP device.
Further, the application supports factor persistence per \S\ref{sec:persistence} as shown in Fig. \ref{fig:persistence}, allowing users to bypass MFA on trusted devices with no reduction in the brute-force resistance of stored MFCHF hashes if those trusted devices are secure.

Finally, the demo application includes two secondary hashes for the purpose of factor recovery, as described in \S\ref{sec:recovery}. Specifically, a HOTP + recovery code hash is stored for password recovery, and a password + recovery code hash is stored for HOTP recovery. Images depicting these features of the demo application are provided in \S\ref{app:appphotos}.

\subsection{Entropy \& Security}
\label{sec:pocentropy}

\begin{figure}[H]
\centering
\includegraphics[width=\linewidth]{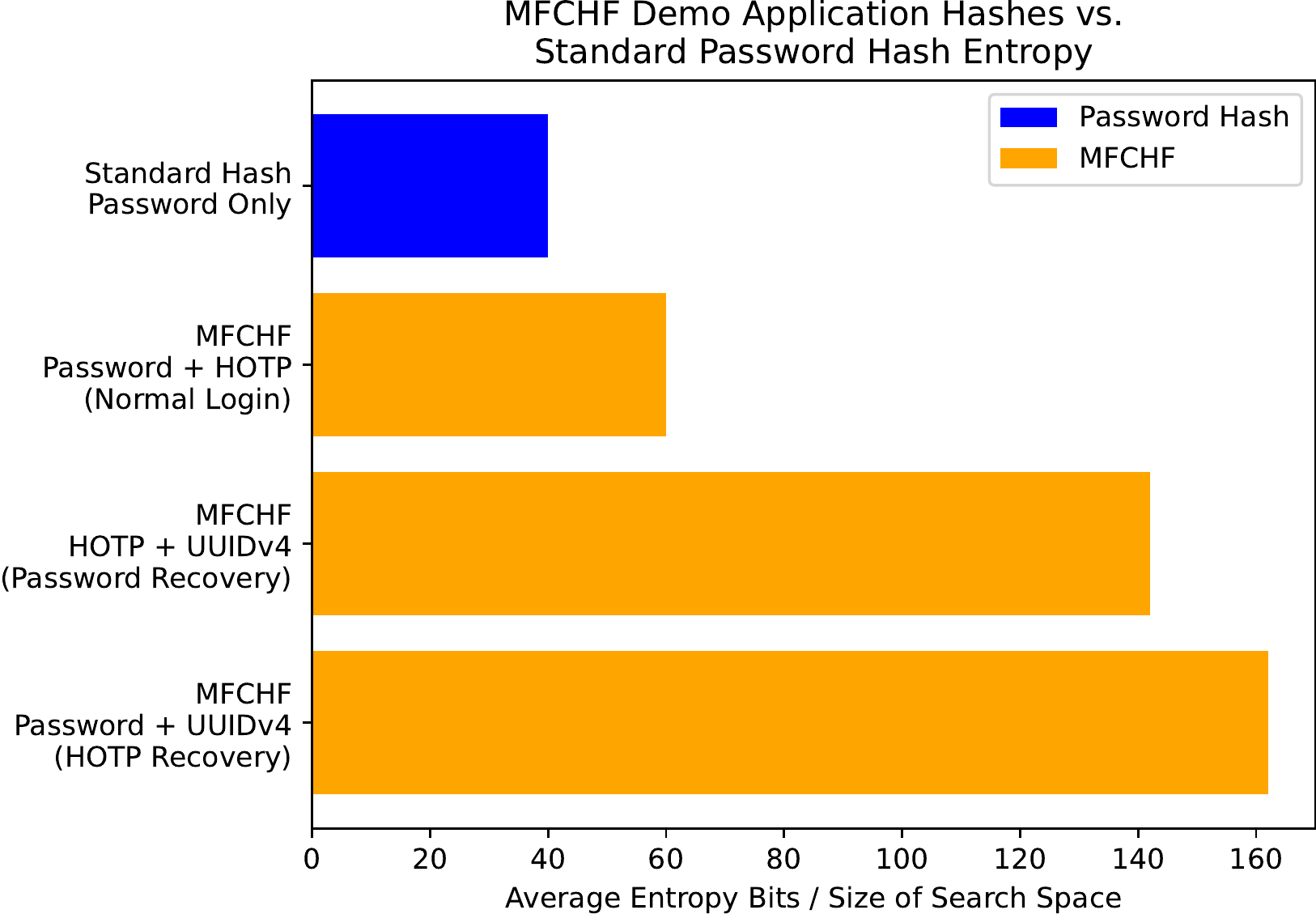}
\caption{\label{fig:entropy}Average brute-force entropy of hashes used in MFCHF proof-of-concept application.}
\end{figure}

To summarize, the demo application uses three distinct multi-factor credential hashes to achieve the desired functionality: a password + HOTP hash for standard logins, a HOTP + recovery code hash for password recovery, and a password + recovery code hash for HOTP recovery. All three of these MFCHF hashes greatly exceed the entropy of a standard password hash ($\approx 40$ bits \cite{florencio_large_2006}), as shown in Fig. \ref{fig:entropy}. Therefore, the weakest link in this application is the password-plus-HOTP hash used for normal authentication, which is still $2^{20}$ (or about one million) times harder to brute-force attack than a standard password hash.

In addition to providing greater asymmetric resistance to brute-force attacks, the absence of HOTP secrets (or recovery codes) stored in plaintext anywhere in the application greatly increases the probability of these secondary factors remaining operational and secure in the interim period between the occurrence of a breach and its detection.

\subsection{Summary}

We present the fully-featured web application demo of this section to illustrate that MFCHF is concretely practical and suitable for real-world deployment. As described above, the scheme used in this proof of concept provides a 1,000,000x increase in asymmetric brute-force attack difficulty and increased post-breach security.
Furthermore, these security advantages implicate very little impact on usability if MFA was already in use, with full backward compatibility with existing HOTP applications. As we have demonstrated, common usability features such as factor persistence, factor recovery, and HOTP validation windows are compatible with MFCHF without reducing its added security. What follows is a general evaluation of the performance and security of the proposed schemes.

%% file: 100-Evaluation.tex
\clearpage

\section{Evaluation}
\label{sec:evaluation}

\vspace{-0.4em}

Although the demo application of \S\ref{sec:pocapp} abstractly addresses the practicality of MFCHF in a realistic setting, we further performed experiments to evaluate the performance of MFCHF in three important aspects. First, we benchmarked the computational and storage overhead of MFCHF over existing password hash functions to highlight its efficiency. Next, we evaluated the theoretical increase in brute-force search space provided by each MFCHF construction. Finally, we performed real brute-force attacks against a variety of schemes to demonstrate the security of MFCHF against a realistic adversary.

\subsection{Performance}
\label{sec:perf}

\vspace{-0.4em}

\begin{figure}[H]
\centering
\includegraphics[width=\linewidth]{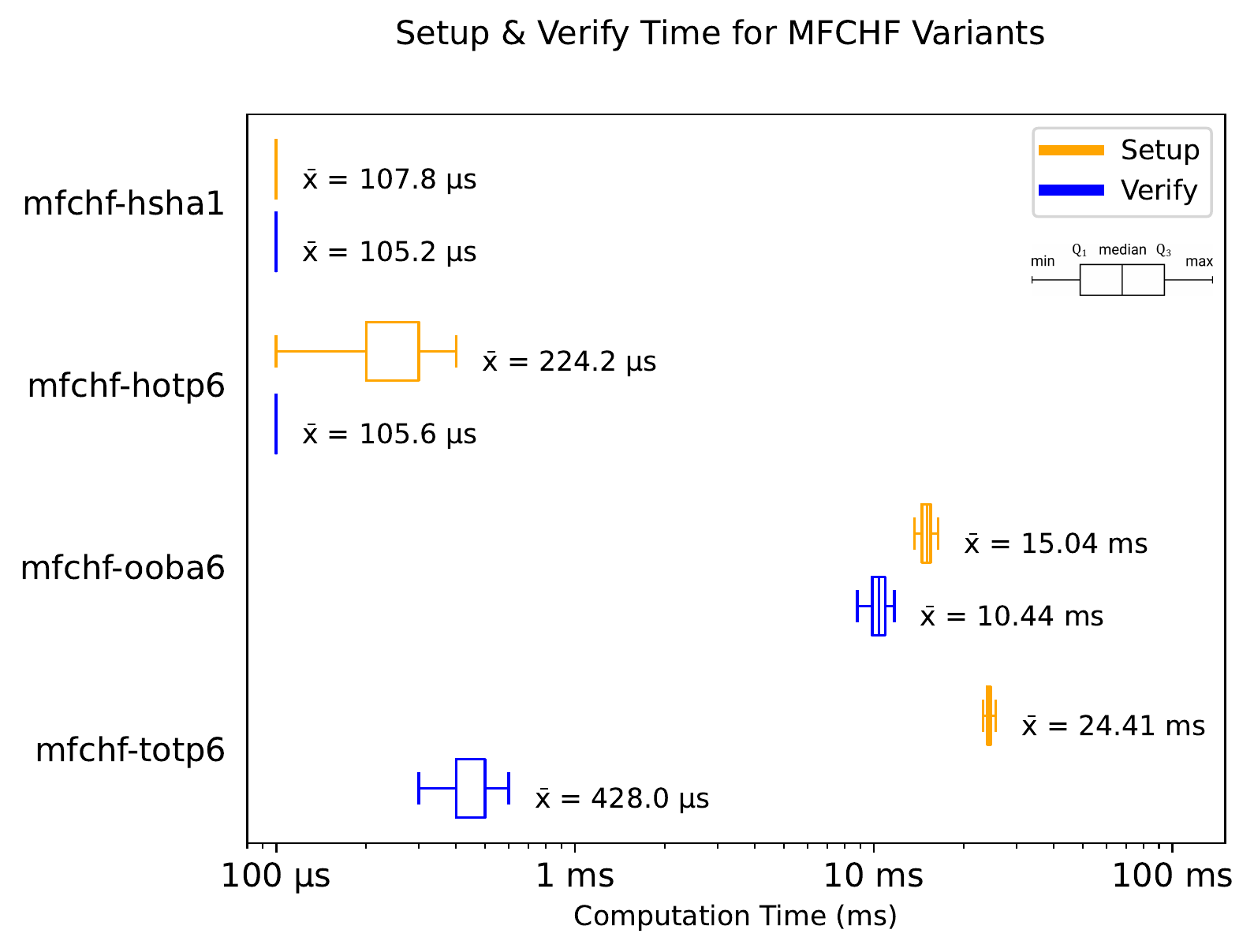}
\caption{\label{fig:overhead}Overhead of MFCHF functions (box plot).}
\end{figure}

To benchmark the performance of MFCHF, we ran the JavaScript library used for the proof-of-concept application of \S\ref{sec:pocapp} using Chrome v106.0.5249.103 on the same benchmarking machine used for all other experiments (see \S\ref{rig}). The $\textsc{Setup}$ and $\textsc{Verify}$ functions for mfchf-hotp6, mfchf-totp6, mfchf-ooba6, and mfchf-hsha1 were run 100 times each to measure the mean overhead of each function. To isolate the computational overhead of MFCHF from that of the underlying hashing mechanism, SHA256 was used as the hash function in all cases. For TOTP, a window of 2,920 was used. The results are shown in Fig. \ref{fig:overhead}.

Another aspect worth evaluating is the space required to store each hash. A standard Argon2 hash requires about 77 bytes. By comparison, the space required increases to 131 bytes for mfchf-hotp6 and mfchf-hsha1 and 353 bytes for mfchf-ooba6. The absolute size of these hashes remains small enough to likely not pose a barrier to adoption, particularly in consideration of the fact that they replace the need to separately store an HMAC secret. However, the space required for mfchf-totp6 could vary from 7 to 219~kb depending on the size of window used.

\subsection{Entropy}

\vspace{-0.4em}

Next, we briefly evaluate the theoretical increase in input entropy (i.e., brute-force search space) of MFCHF over standard password hashes. For the HOTP and TOTP variants of MFCHF, adversaries must evaluate all $10^6$ possible target values for each attempted password. Because OOBA can use an alphanumeric OTP, adversaries must evaluate all $36^6$ possible target values for each attempted password. Finally, for the HMAC-SHA1 (i.e., YubiKey) variant, the HMAC key is included in the hash, which can have $2^{160}$ possible values. Overall, the theoretical entropy gained by mfchf-hotp6, mfchf-totp6, mfchf-ooba6, and mfchf-hsha1 is 20, 20, 31, and 160 bits, respectively.


\subsection{Brute-Force Resistance}
\label{sec:bfr}

\vspace{-0.4em}

\begin{figure}[H]
\centering
\includegraphics[width=\linewidth]{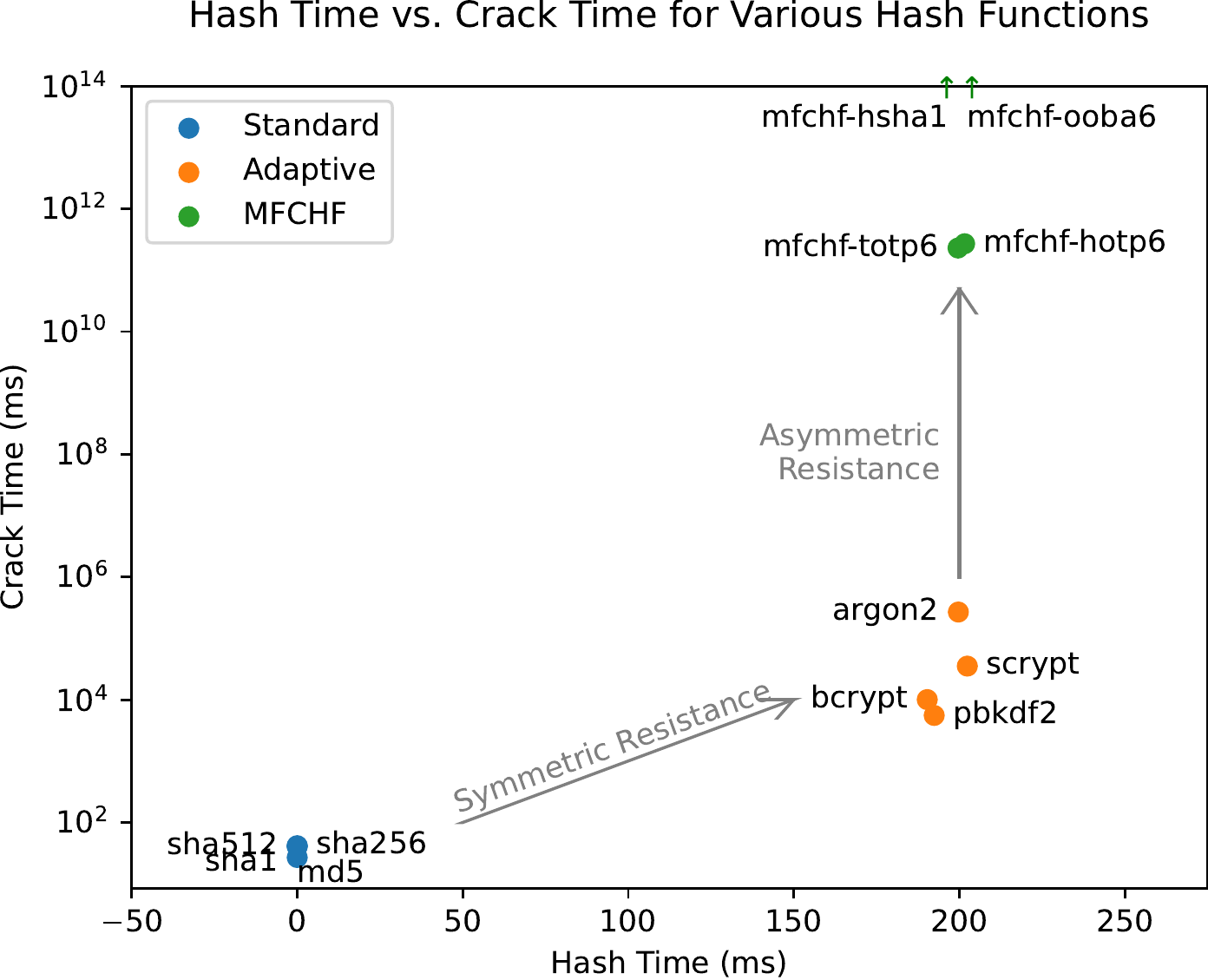}
\caption{\label{fig:eval}Hash and crack time for various hash types.}
\end{figure}

While the theoretical increase in search space of MFCHF strongly indicates an improvement in brute-force difficulty, we also performed an experiment to validate the increased brute-force resistance of MFCHF. We first created 100 salted password hashes for each of several standard (MD5, SHA1, SHA256, and SHA512) and adaptive (PBKDF2, bcrypt, scrypt, and Argon2) hashing schemes. The hashes were chosen from a dictionary of the 10,000 most common passwords, and cost parameters for adaptive hashes were configured to take 200~ms to compute. Next, we used John the Ripper (for Argon2) or Hashcat (for all others) to crack each password via an exhaustive search of the dictionary. The hash and crack times for each hash type are shown in Fig. \ref{fig:eval}, with the full results in \S\ref{sec:evalresults}.

Next, we evaluated the brute force time of four types of MFCHF hashes (mfchf-hotp6, mfchf-totp6, mfchf-ooba6, and mfchf-hsha1) with Argon2id as the underlying hash function. Given that each of these hashes would take well over a year to crack using the above method, we instead let John the Ripper run for exactly 24 hours for each of the four hashes and examined the percentage of the search space exhausted in that time. We found that 0.016\% and 0.019\% of the search space were exhausted after 24 hours for mfchf-hotp6 and mfchf-totp6, respectively. Given that cracking a hash would require attempting 50\% of the search space on average, we expect the average crack times for mfchf-hotp6 and mfchf-totp6 to be about 8.6 years and 7.2 years, respectively. For mfchf-ooba6 and mfchf-hsha1, John the Ripper reported 0.000\% progress after 24 hours. Based on the relative entropy, we expect these hashes would take thousands of years to crack. These MFCHF results are also included in Fig. \ref{fig:eval} and in \S\ref{sec:evalresults}.

\subsection{Discussion}

In \S\ref{sec:hashing}, we introduced the notions of \textit{symmetric resistance}, where brute-force attack difficulty increases proportionally with an increase in verification time, and \textit{asymmetric resistance}, where brute-force attack difficulty is disproportionately improved relative to verification time for legitimate users. As demonstrated by our evaluation, MFCHF provides a robust improvement in asymmetric brute-force attack resistance: with the same $200$~ms target as bcrypt, scrypt, PBKDF2, and Argon2, MFCHF provides a $10^6$ to $10^{48}$ times increase in the difficulty of cracking hashed credentials. It does so by leveraging the entropy of passwords and multi-factor authentication together within a single multi-factor credential hash, rather than simply increasing the computational difficulty of the hash function, as is seen in most adaptive schemes.

Indeed, the practical effect of even the weakest MFCHF variants (HOTP/TOTP) far exceeds what one might intuitively expect given the seemingly small amount of added entropy (20~bits). While the average crack time for this setup without MFCHF was just 4.5~minutes, using MFCHF would increase the time to crack such a hash to over 7~years with no impact on the verification time for a legitimate user. This indicates that the current landscape of adaptive password hashing, with about 40~bits of entropy in the average password and 200~ms of latency tolerance, exists precisely at an optimality point for MFCHF to have a dramatic impact on the feasibility of a brute-force attack.

Interestingly, our empirical results do not represent a simple linear improvement in brute-force resistance relative to the added entropy. For example, most adaptive hash functions were susceptible to highly-parallel GPU-based attacks, significantly reducing their security in practice. Indeed, our chosen dictionary size of 10,000 was sufficiently large that almost all 10,496 GPU cores could be deployed in parallel to attack adaptive hashes (while still being small enough to feasibly evaluate within a short period of time). Increasing the dictionary size further could increase the attack difficulty across all hash types, but should not affect the relative advantage of MFCHF.

Finally, the computational overhead of some MFCHF variants required a reduction in underlying cost parameters to maintain a 200~ms latency target, which could reduce the added resistance in practice. For instance, while the overhead of all MFCHF functions was relatively small in relation to the 200~ms target, the higher overhead of mfchf-totp6 compared to mfchf-hotp6 resulted in mfchf-hotp6 providing more brute-force resistance than mfchf-totp6, despite both functions theoretically providing the same amount (20~bits) of additional entropy. 

%% file: 150-Security-Analysis.tex
\section{Security Analysis}
\label{sec:security}

Having now empirically demonstrated the asymmetric resistance of MFCHF, we next provide brief arguments for why the schemes of \S\ref{sec:mfchf} satisfy the security properties of \textit{correctness}, \textit{safety}, and \textit{asymmetric resistance}, as defined in \S\ref{sec:goals}. While a formal proof framework for multi-factor key derivation exists \cite{krawczyk, mfkdf}, there is no suitable equivalent for multi-factor credential hashing. Our arguments are instead based on a semi-formal reduction to the security properties of the underlying hash functions.

\subsection{Primitives}
The MFCHF constructions of \S\ref{sec:mfchf} use an adaptive password hash function such as Argon2 ($\mathsf{H_1}$), and a standard cryptographic hash function such as SHA-256 ($\mathsf{H_2}$). For both hash functions, the following properties are required:

\begin{itemize}[leftmargin=*]
    \item \textit{Determinism} -- For a given input value $m$, $\mathsf{H}(m)$ must always generate the same hash value $h$ with $p=1$.
    \item \textit{Total Pre-Image Resistance} -- Given only an arbitrary hash value $h$, it is hard for an adversary to find any bit of input $m$ such that $\mathsf{H}(m) = h$ except with $p=0.5+negl$.
    \item \textit{Second Pre-Image Resistance} -- Given an input $m_1$, it is hard for an adversary to find another input $m_2$ such that $\mathsf{H}(m_1) = \mathsf{H}(m_2)$ and $m_1 \neq m_2$ except with $p=negl$.
    \item \textit{Uniformly Pseudorandom} -- Given an arbitrary $m$ and $\mathsf{H}(m)=h$, each bit of $h$ is $1$ with $p=0.5$.
    \item \textit{Strict Avalanche Criterion} -- Given $\mathsf{H}(m)=h$, changing one bit of $m$ should change each bit of $h$ with $p=0.5$.
\end{itemize}


\subsection{Correctness}
\label{sec:correctness}
The correctness property of MFCHF requires that when provided with valid witnesses corresponding to a user's established authentication factors and hash, the server outputs $\mathit{accept}$ with $p=1$. For a given login request, there is exactly one valid password and witness.
In each MFCHF construction, the server outputs $\mathit{accept}$ if  $\mathsf{H}(\mathsf{password} \odot \mathsf{target} \odot \mathsf{salt})$ matches a stored hash of the correct values. It directly follows from the \textit{determinism} property of $\mathsf{H}$ that if the correct password and target are provided, $\mathsf{H}$ will generate the same value $h$ (and therefore, the server will output $\mathit{accept}$) with $p=1$.

What remains to be shown is that $\mathsf{target}$ will match the expected value if the correct OTP is provided.
In the case of the HMAC-SHA1 construction, this is true because bitwise XOR with one parameter fixed is an involution ($A \oplus B \oplus B = A)$: $\mathsf{key} \oplus \mathsf{response} \oplus \mathsf{response} = \mathsf{key}$. For all other constructions, this instead holds because of the congruence relation $((A - B)~\%~N + B)~\%~N \equiv A$, which suggests $((A - B)~\%~N + B)~\%~N = A$ when $A \in [0, N)$. Thus, $((\mathsf{target} - \mathsf{otp})~\%~10^6 + \mathsf{otp})~\%~10^6 = \mathsf{target}$.



\subsection{Safety}
The safety property of MFCHF posits that when at least one of a user's factor witnesses is invalid, the server outputs $\mathit{reject}$ except with $p=negl$.
In each MFCHF construction, the server outputs $\mathit{reject}$ if $\mathsf{H}(\mathsf{password} \odot \mathsf{target} \odot \mathsf{salt})$ does not match a stored hash of the correct values.
If an adversary can violate the safety of MFCHF by finding an invalid $\mathsf{password}$ or $\mathsf{target}$ value that causes $\mathsf{H}$ to generate the same value $h$ as the correct credentials (and therefore, the server to output $\mathit{accept}$), then they have also violated the 
\textit{second pre-image resistance} of $\mathsf{H}$ by finding $m_1 \neq m_2$ such that $\mathsf{H}(m_1) = \mathsf{H}(m_2)$.

It remains to be shown that $\mathsf{target}$ will not match the expected value if the incorrect OTP is provided.
To illustrate this, we simply invert the proofs given in \S\ref{sec:correctness}: $A \oplus B \oplus B' \neq A$ if $B \neq B'$ (thus $\mathsf{key}$ is wrong if $\mathsf{response}$ is wrong), and $((A - B)~\%~N + B')~\%~N \not\equiv A$ if $B \not\equiv B'$ (thus $\mathsf{target}$ is wrong if $\mathsf{otp}$ is wrong).

\subsection{Asymmetric Resistance}

Finally, the asymmetric resistance property of MFCHF suggests that an MFCHF hash should be significantly harder to crack than an adaptive password hash with the same fixed verification time. Specifically, the brute-force difficulty is exponential with respect to the total entropy of the input space, so an MFCHF hash with a password and a 6-digit OTP should be $10^6$ times harder to crack than a comparable password hash.

Given $\mathsf{H}(m) = h$, the \textit{avalanche effect} of $\mathsf{H}$ suggests that if $m$ is changed slightly (e.g, flipping a single bit), $h$ changes significantly (e.g., half the bits flip).
Now given $\mathsf{H}(\mathsf{password} \odot \mathsf{target} \odot \mathsf{salt}) = h$, the asymmetric resistance of the MFCHF hash $h$ derives from the \textit{avalanche effect} of $\mathsf{H}$, as any incorrect bit of $\{\mathsf{password} \odot \mathsf{target}\}$ will cause $h$ to be completely incorrect. Furthermore, the \textit{total pre-image resistance} of $\mathsf{H}$ precludes an adversary from reversing any part of $h$ to determine which bits are incorrect. Given $\lambda_1$ bits of entropy in $\mathsf{password}$ and $\lambda_2$ bits of entropy in $\mathsf{target}$, all $2^{\lambda_1 + \lambda_2}$ possible values must be exhaustively searched. The effect of adding a 6-digit ($\approx 20$-bit) OTP is thus to increase the brute-force difficulty of $h$ by a factor $\approx 2^{20}$ or $10^6$ for the same verification time, resulting in asymmetric resistance.

Lastly, it remains to be argued that the values stored alongside the hash and salt in MFCHF do not weaken or reveal secret information about the underlying factors:

\begin{itemize}[leftmargin=*]
    \item For the $\mathsf{blind}$ value, this is the case because assuming the output of $\mathsf{H}$ is \textit{uniformly pseudorandom}, it can be used as a one-time pad (Vernam cipher) to hide the value of $\mathsf{key}$ with information-theoretic security.
    \item For the $\mathsf{diff}$ value, this holds because of the security of modular arithmetic (rings); a pseudorandom $\mathsf{otp}$ can be used to hide $\mathsf{target}$ with information-theoretic security.
    \item For HMAC-SHA1, the $\mathsf{challenge}$ value is chosen uniformly randomly in $[0, 2^{160})$. If HMAC-SHA1 is secure, $\mathsf{challenge}$ does not reveal $\mathsf{key}$ or $\mathsf{response}$.
    \item For OOBA, the values of $\mathsf{ct}$ and $\mathsf{pk}$ do not reveal $\mathsf{otp}$ if the public-key encryption method is IND-CCA secure.
\end{itemize}

\section{Limitations}
\label{sec:limitations}

The first and most obvious limitation of MFCHF is that it requires the use of one of the supported multi-factor authentication methods. While MFA is increasingly widely supported by a variety of online services, the rate of voluntary adoption by end users has remained slow.
Thus, it must be emphasized that our solution is chiefly targeted at services that already implement one of the supported authentication factors, and will still only benefit the subset of users who choose to enable MFA.

The above limitation is further evident in the recovery setup of \S\ref{sec:recovery}, which requires three independent factors. It is already often the case that an online service supporting one MFA factor (e.g., TOTP) will also allow for a lost password to be recovered via another factor (e.g., Email OOBA). Still, the need to establish multiple factors for recovery may pose a barrier to the adoption of this method.

Another limitation is the need for a PKI in the OOBA variant.
For email, S/MIME \cite{rfc3850} is a widely accepted protocol that provides a PKI that can be used for this purpose and is supported by the majority of modern email software. This can be extended to SMS via the email-to-SMS gateway services offered by most major phone carriers \cite{sms_gateway}.
However, not all email providers support S/MIME, and not all phone carriers provide an email-to-SMS gateway.
Therefore, OOBA is the only MFCHF variant that relies on features without universal support, and some users, such as those using legacy mail clients, will not be able to benefit from the OOBA construction.

With respect to the TOTP construction, one limitation is the need to choose a large enough window of offsets that a user does not lose access to their account if inactive for long periods.
The need to calculate a large number of offset values is the reason TOTP has a higher setup overhead than all other MFCHF variants in \S\ref{sec:perf}.
As a mitigating factor, we note that the calculation of offsets in the TOTP construction occurs after the $\mathit{accept}$ or $\mathit{reject}$ determination has already been made (see Alg. \ref{alg:mfchf_totp6} in \S\ref{app:algs}). Thus, a service could opt to update offset values asynchronously in the background after authenticating a user.

Finally, MFCHF fundamentally requires that all authentication factors be verified simultaneously, rather than being verified sequentially, as is often currently the case. While this provides a marked security improvement by preventing each factor from being individually attacked, it could prove frustrating for a user who fails to authenticate and is unsure which factor is incorrect. In the future, a method that statistically determines which factor is wrong could be used without losing much entropy in the process.

\section{Failure Modes}

Per our threat model (\S\ref{sec:threat-model}), the failure of an MFCHF hash occurs if an adversary is able to reverse the hash to obtain any of the underlying authentication factors.
Our security analysis argues that the only ways to defeat an MFCHF hash are (1) to compromise the underlying factors or (2) to perform a brute-force attack.
Still, there are several scenarios in which either may occur, even if MFCHF is correctly implemented with secure primitives. 

Clearly, if a combination of factors with insufficient total entropy (such as a 6-digit OOBA factor and a 6-digit HOTP factor) is chosen, brute-force attacks against the entire input space may still be feasible.
What is perhaps less obvious is that even if a single factor is weak, the entire hash may be attacked. For example, though an MFCHF hash combining a password and HOTP factor may itself be infeasible to crack, the compromise of the password factor may allow the remaining factor to be defeated by brute force. However, in such a scenario, the marginal value of compromising the second factor is limited, as HMAC secrets, unlike passwords, are rarely shared across accounts.
Still, the use of password strength requirements and compromised credential checking remains essential.

Additionally, the ``total data breach'' threat model described in \S\ref{sec:threat-model} does not consider the risk of an ongoing threat like undetected malware on the authentication server. In the scenario where the server remains actively compromised for long periods of time, authentication secrets can be stolen during the ephemeral period in which they are decrypted upon user login. While still an improvement over the typical method of storing HMAC, HOTP, and TOTP keys in plaintext, this constitutes another potential failure mode in which factors become compromised.

Finally, it is expected that a system implementing MFCHF is still only as secure as its underlying factors, and these factors must be properly managed and protected. For example, if an SMS device is vulnerable to a SIM-swapping attack, then an mfchf-ooba6 hash utilizing that device would be susceptible to compromise.

%% file: 200-Related-Work.tex
\section{Related Work}
\label{sec:related}

Within the field of password hashing, there are no known works describing hash functions that incorporate common authentication factors like HOTP, TOTP, or YubiKey (HMAC-SHA1) into the hashing process so as to form a multi-factor credential hash. Instead, the vast majority of research into attack-resistant password hashing has focused on symmetric brute-force resistance, including works such as PBKDF2 \cite{pbkdf2}, bcrypt \cite{bcrypt}, scrypt \cite{scrypt}, yescrypt \cite{yescrypt}, and Argon2 \cite{argon2} that provide various degrees of hardware resistance. Other works, such as Catena \cite{catena} and Lyra2 \cite{lyra2}, emphasize resistance to side-channel attacks. Finally, delegable password hashing functions like Makwa \cite{makwa} are designed to allow clients to outsource computation power for password hashing to untrusted third parties, thereby potentially increasing the computational power available to the client.

The relative resistance of password hashing functions to various hardware-enabled adversaries (e.g., CPU, GPU, or ASIC-based threats) has been the subject of benchmarking experiments \cite{phc_benchmarks}. However, we did not find any other studies surveying the effect of these functions on the actual rate of password disclosure across real data breaches.

With respect to multi-factor credential hashing, the ``factor constructions'' of MFKDF \cite{mfkdf} were the main inspiration for the techniques of this paper. In MFKDF, these techniques are applied on the client side to derive a key for the purpose of end-to-end encryption, rather than being used on the server side for credential hashing. In particular, the authentication technique suggested by MFKDF requires the establishment of a shared key, which, unlike the techniques of this paper, would immediately fail to provide secure authentication after a data breach (although stored secrets would remain secure). Furthermore, the three-part scheme of MFKDF ($\textsc{Setup}$, $\textsc{Derive}$, and $\textsc{Update}$) are simplified into a more efficient two-part scheme ($\textsc{Setup}$ and $\textsc{Verify}$) in this work.

Prior to MFKDF, several works have proposed two-factor key derivation based specifically on a password and a YubiKey hardware token \cite{2fe, parker_enhancing}. Once again, these works focus on client-side deployment for an end-to-end encryption use case. Other than MFKDF, there are no known prior works utilizing HOTP or TOTP for key derivation, let alone credential hashing. While other works have addressed using these factors for the related problem of Multi-Factor Authenticated Key Exchange (MFAKE) \cite{mfake, mfake2, mfake3}, this approach aims to establish a shared key in the presence of man-in-the-middle adversaries and does not provide brute-force resistance or protection against a compromised server. 

%% file: 210-Future-Work.tex
\section{Future Work}
\label{sec:future}

\subsection{Factors}
Our focus in this paper was on those factors for which constructions were provided in MFKDF, namely static factors (e.g., passwords and recovery codes), HOTP, TOTP, HMAC-SHA1 (YubiKey), and out-of-band authentication (OOBA) factors such as SMS and email. While support for arbitrary factors based on MPC and trusted hardware are possible, they were not included in this paper due complicating the security model, but can be included in future work. With respect to TOTP, future work should emphasize reducing the size of data stored from the 219 kb currently required. Finally, future research should investigate the incorporation of factors not currently supported by MFKDF, such as biometrics, geolocation, device fingerprinting, behavioral authentication, and OIDC.

\subsection{Features}
Although the account recovery feature described in this paper provides sufficient functionality for the vast majority of systems, future work could take advantage of the threshold and policy-based MFKDF variants for use in systems with more advanced authentication policies. Furthermore, future iterations of MFCHF should ideally facilitate progressive deployment on systems that currently use password hashing without requiring all passwords to be reset. Lastly, future research could analyze the usability impact of MFCHF, specifically with regard to the requirement to provide and validate all authentication factors simultaneously. Mechanisms for mitigating this impact are also worth investigating, though this requirement may, in fact, be an inevitable trade-off of the MFCHF approach.
Specifically, as mentioned in \S\ref{sec:limitations}, we would love to see a feature that relinquishes a small amount of entropy to probabilistically ``hint'' the user which factor is wrong in the event that a login attempt fails to validate.

\subsection{Applications}
While we focus on the server-side use of MFCHF in the client-server setting, future work should explore the use of MFCHF in other applications, such as for user authentication within operating systems. Crypt, the predecessor to Bcrypt, was originally developed for password hashing in Unix, so as to enable password-based authentication without storing passwords in plaintext. While using authentication factors like HOTP within operating systems would typically not have been feasible due to the need to store an HMAC key, MFCHF can enable the secure use of such factors for operating system authentication. The potential use of MFCHF for authentication in networks and decentralized systems should also be investigated.

\eject

%% file: 220-Conclusion.tex
\section{Conclusion}
With both the volume and severity of major information security incidents continuing to experience exponential growth, password hashing has played an important role in reducing the rate of credential disclosure (and thus potential liability) resulting from a data breach. Indeed, our analysis of over 4,000 actual data breaches containing hashed credentials clearly demonstrates that the use of salted and adaptive password hashing has a significant effect on the rate of password disclosure not just in theory, but also in practice. Still, while the majority of research in this area has emphasized the use of adaptive hash functions for symmetric brute-force attack resistance, the low tolerance of users for added latency places an upper limit on the applicability of this technique in practice.

Coinciding with the rise in large-scale data breaches is the increased use of multi-factor authentication to combat the threat of credential stuffing. We thus set out to create a multi-factor credential hashing function that utilizes the additional entropy provided by multi-factor authentication to provide asymmetric resistance to brute-force attacks. Our scheme emphasizes usability by supporting common features like account recovery, desynchronization windows, and factor persistence, while maintaining compatibility with popular, unmodified authentication factors like HOTP, TOTP, and YubiKey that are already in use.

To demonstrate the practicality of MFCHF as a drop-in replacement for password hashing, we produced a full implementation of MFCHF using HOTP and evaluated its deployment within a typical full-stack web application. Our evaluation shows MFCHF to be dramatically harder to brute-force attack, while having a low computational overhead with negligible impact on setup and verification time for a legitimate user. Overall, in systems where MFA is already in use, MFCHF provides a compelling value proposition over traditional password hashing by improving brute-force resistance and post-breach security with limited impact on usability or performance.

%% file: 950-Appendix-Demo.tex
\section{Demo Application}
\label{app:appphotos}

\begin{figure}[H]
    \centering
    
    \begin{subfigure}[b]{0.49\linewidth}
        \centering
        \includegraphics[width=\textwidth]{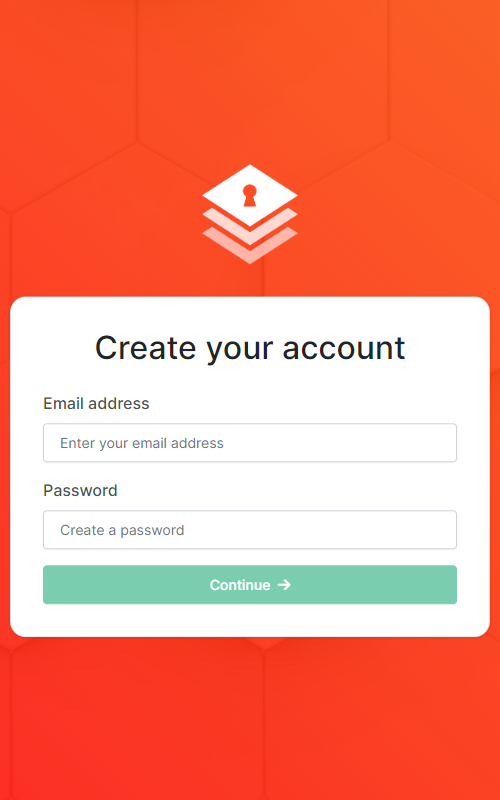}
        \caption{Registration}
        \label{fig:registration}
    \end{subfigure}
    \hfill
    \begin{subfigure}[b]{0.49\linewidth}
        \centering
        \includegraphics[width=\textwidth]{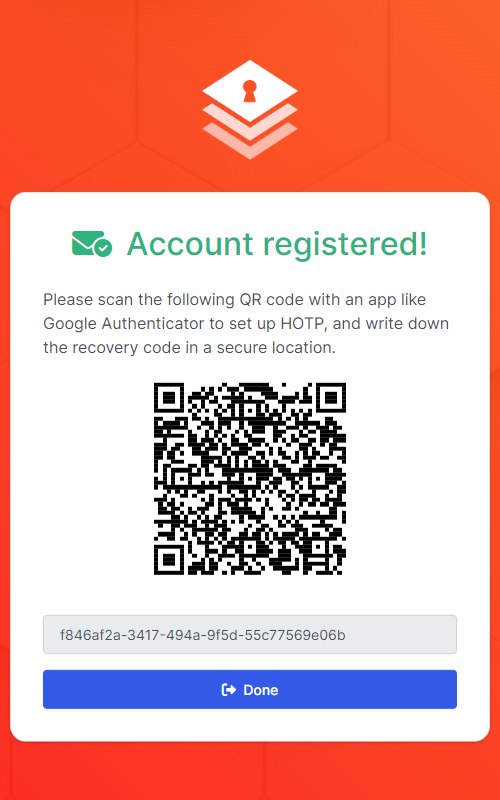}
        \caption{HOTP Setup}
        \label{fig:hotpsetup}
    \end{subfigure}
    
    \vspace{0.1 \linewidth}
    
    \begin{subfigure}[b]{0.49\linewidth}
        \centering
        \includegraphics[width=\textwidth]{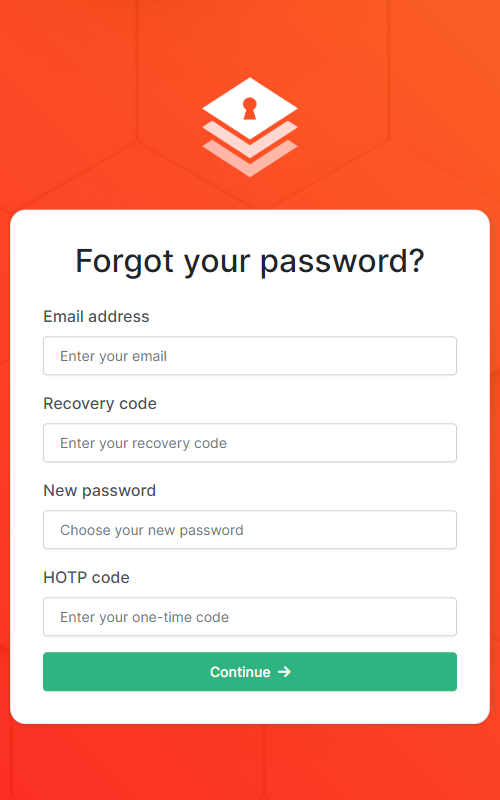}
        \caption{Password Recovery}
        \label{fig:passwordrecovery}
    \end{subfigure}
    \hfill
    \begin{subfigure}[b]{0.49\linewidth}
        \centering
        \includegraphics[width=\textwidth]{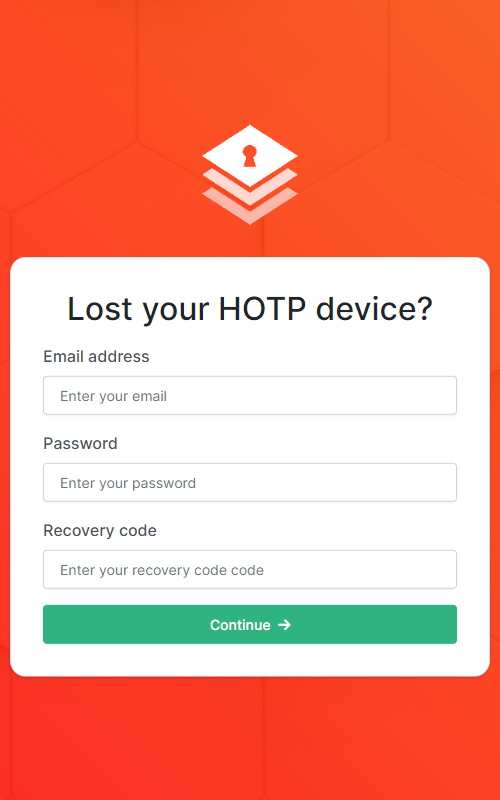}
        \caption{HOTP Recovery}
        \label{fig:hotprecovery}
    \end{subfigure}
    
    \vspace{0.1 \linewidth}
    
    \begin{subfigure}[b]{0.49\linewidth}
        \centering
        \includegraphics[width=\textwidth]{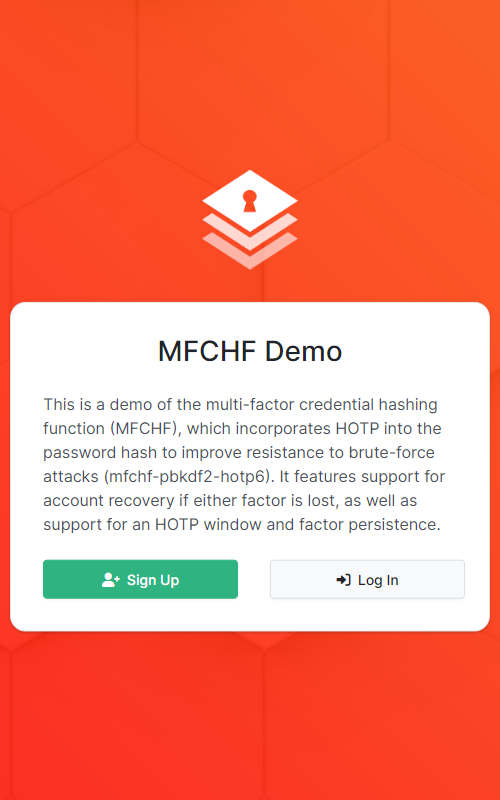}
        \caption{Home Screen}
        \label{fig:home}
    \end{subfigure}
    \hfill
    \begin{subfigure}[b]{0.49\linewidth}
        \centering
        \includegraphics[width=\textwidth]{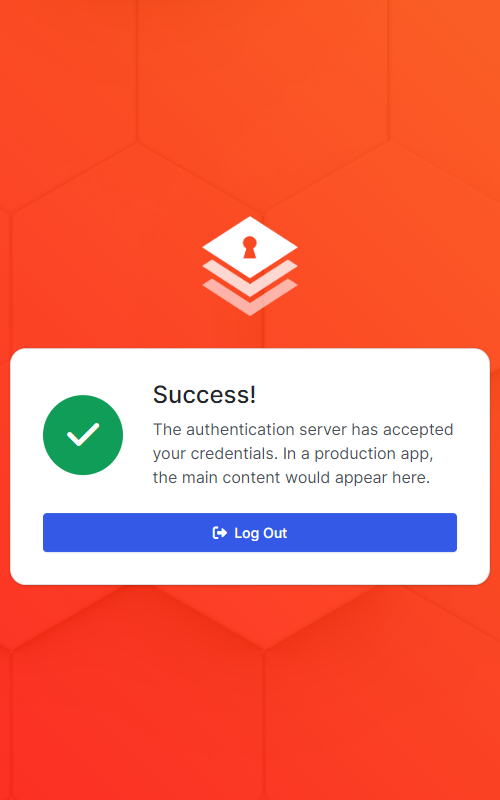}
        \caption{Internal Application}
        \label{fig:internal}
    \end{subfigure}

    \caption{Authentication screens for demo application.}
    \label{fig:centralizedappx}
\end{figure}

%% file: 910-Appendix-Specs.tex
\clearpage

\onecolumn

\section{Specifications}
\label{rig}

\vspace{-0.3em}

\noindent For reproducibility, we include here the exact specifications of the device used for all benchmarking and evaluation throughout this paper. We expect the general trends observed in the results to hold regardless of the hardware used.

\begin{itemize}[leftmargin=*]
    \itemsep 0em
    \item \textbf{CPU}: AMD Ryzen 9 5950X (16-core, 3.40 GHz)
    \item \textbf{GPU}: NVIDIA GeForce RTX 3090 (10496-core, 24.0 GB)
    \item \textbf{RAM}: 64.0 GB (2x32, 3200 MHz)
    \item \textbf{SSD}: 2.0 TB NVMe M.2 (PCIe Gen3x4, 3400 MB/s)
\end{itemize}

%% file: 920-Appendix-Table.tex
\section{Hashmob Results}
\label{fulldata}

\vspace{-0.3em}

\begin{table}[H]
\resizebox{\textwidth}{!}{%
\begin{tabular}{|l|l|l|l|l|}
\hline
\textbf{Hash Type} & \textbf{Hashcat Mode} & \textbf{\# of Data Breaches} & \textbf{\% of Hashes Cracked} & \textbf{Mean Crack Time} \\ \hline
MD5 & 0 & 1553 & 71.13\% ($\sigma$=35.57\%) & 7.6d ($\sigma$=36.1d) \\ \hline
vBulletin \textless v3.8.5 & 2611 & 406 & 63.54\% ($\sigma$=27.65\%) & 8.0d ($\sigma$=38.5d) \\ \hline
vBulletin \textgreater{}= v3.8.5 & 2711 & 379 & 57.49\% ($\sigma$=27.67\%) & 0.1d ($\sigma$=48.9d) \\ \hline
bcrypt & 3200 & 283 & 12.37\% ($\sigma$=19.86\%) & 71.2d ($\sigma$=89.3d) \\ \hline
phpass & 400 & 214 & 32.52\% ($\sigma$=30.02\%) & 1205.2d ($\sigma$=15189.7d) \\ \hline
SHA1 & 100 & 198 & 74.01\% ($\sigma$=36.47\%) & 7.5d ($\sigma$=37.7d) \\ \hline
MyBB 1.2+ IPB2+ & 2811 & 162 & 63.96\% ($\sigma$=23.72\%) & 5.3d ($\sigma$=38.5d) \\ \hline
AuthMe sha256 & 20711 & 158 & 83.35\% ($\sigma$=16.20\%) & 8.1d ($\sigma$=17.1d) \\ \hline
Django (PBKDF2-SHA256) & 10000 & 85 & 15.84\% ($\sigma$=18.67\%) & 54.5d ($\sigma$=74.0d) \\ \hline
md5(md5(\$pass)) & 2600 & 81 & 71.24\% ($\sigma$=30.94\%) & 4.2d ($\sigma$=20.4d) \\ \hline
osCommerce xt:Commerce & 21 & 74 & 73.21\% ($\sigma$=32.59\%) & 13.1d ($\sigma$=44.5d) \\ \hline
MySQL4.1/MySQL5 & 300 & 73 & 76.82\% ($\sigma$=37.97\%) & 2.3d ($\sigma$=17.9d) \\ \hline
bcrypt(md5(\$pass)) & 25600 & 54 & 24.46\% ($\sigma$=20.51\%) & 351.9d ($\sigma$=72.7d) \\ \hline
md5($salt.$pass) & 20 & 48 & 67.73\% ($\sigma$=29.34\%) & 10.9d ($\sigma$=80.3d) \\ \hline
WPA-PBKDF2-PMKID+EAPOL & 22000 & 45 & 5.48\% ($\sigma$=13.96\%) & 75.7d ($\sigma$=85.6d) \\ \hline
SHA2-256 & 1400 & 40 & 56.37\% ($\sigma$=39.57\%) & 16.5d ($\sigma$=34.3d) \\ \hline
md5crypt & 500 & 38 & 33.09\% ($\sigma$=29.45\%) & 14.0d ($\sigma$=37.1d) \\ \hline
md5($pass.$salt) & 10 & 36 & 61.83\% ($\sigma$=28.35\%) & 25.6d ($\sigma$=53.6d) \\ \hline
Joomla \textless 2.5.18 & 11 & 32 & 60.54\% ($\sigma$=36.14\%) & 36.5d ($\sigma$=47.6d) \\ \hline
NTLM & 1000 & 31 & 36.94\% ($\sigma$=36.45\%) & 23.2d ($\sigma$=51.1d) \\ \hline
SMF \textgreater v1.1 & 121 & 27 & 57.49\% ($\sigma$=30.67\%) & 59.4d ($\sigma$=82.0d) \\ \hline
MySQL323 & 200 & 26 & 82.18\% ($\sigma$=30.85\%) & 13.8d ($\sigma$=39.5d) \\ \hline
OpenCart & 13900 & 25 & 37.61\% ($\sigma$=21.86\%) & 36.9d ($\sigma$=56.7d) \\ \hline
sha1($salt.$pass) & 120 & 24 & 63.74\% ($\sigma$=25.05\%) & 42.0d ($\sigma$=70.3d) \\ \hline
nsldaps SSHA-1(Base64) & 111 & 22 & 16.32\% ($\sigma$=7.30\%) & 143.0d ($\sigma$=38.4d) \\ \hline
Drupal7 & 7900 & 20 & 15.43\% ($\sigma$=29.53\%) & 144.3d ($\sigma$=162.3d) \\ \hline
Django (SHA-1) & 124 & 19 & 76.37\% ($\sigma$=20.46\%) & 17.6d ($\sigma$=10.9d) \\ \hline
sha256($salt.$pass) & 1420 & 18 & 58.60\% ($\sigma$=27.83\%) & 33.9d ($\sigma$=32.2d) \\ \hline
SHA2-512 & 1700 & 18 & 51.91\% ($\sigma$=35.59\%) & 28.2d ($\sigma$=68.4d) \\ \hline
sha512($salt.$pass) & 1720 & 17 & 31.82\% ($\sigma$=26.60\%) & 82.9d ($\sigma$=15.5d) \\ \hline
descrypt & 1500 & 15 & 56.91\% ($\sigma$=40.26\%) & 20.9d ($\sigma$=56.7d) \\ \hline
Ruby on Rails Restful Auth & 27200 & 14 & 67.07\% ($\sigma$=23.94\%) & 57.6d ($\sigma$=71.1d) \\ \hline
sha1($pass.$salt) & 110 & 12 & 40.88\% ($\sigma$=39.77\%) & 40.8d ($\sigma$=106.7d) \\ \hline
sha256($pass.$salt) & 1410 & 12 & 40.00\% ($\sigma$=31.05\%) & 38.8d ($\sigma$=58.0d) \\ \hline
\end{tabular}%
}
\caption{Summary of 4,259 data breaches on HashMob, excluding formats with $\leq 10$ data breaches or 0 cracks.}
\label{tab:hashmob}
\end{table}

%% file: 930-Appendix-Results.tex
\section{Evaluation Results}
\label{sec:evalresults}

\vspace{-0.3em}

\begin{table}[H]
\resizebox{\textwidth}{!}{%
\begin{tabular}{|l|l|l|l|l|}
\hline
\textbf{Hash Type} & \textbf{Attack Mode} & \textbf{Cost Parameters} & \textbf{Mean Verification Time} & \textbf{Mean Crack Time} \\ \hline
MD5 & Hashcat 10 & n/a & 16.9 $\mu$s & 26.91 ms \\ \hline
SHA1 & Hashcat 110 & n/a & 21.1 $\mu$s & 41.07 ms \\ \hline
SHA256 & Hashcat 1410 & n/a & 10.6 $\mu$s & 41.85 ms \\ \hline
SHA512 & Hashcat 1710 & n/a & 10.2 $\mu$s & 40.93 ms \\ \hline
PBKDF2-SHA2 & Hashcat 20300 & n=1000000 & 192.4 ms & 5552.75 ms ($\approx$ 5.5 s) \\ \hline
Bcrypt & Hashcat 3200 & c=12 & 190.3 ms & 10071.5 ms ($\approx$ 10 s) \\ \hline
Scrypt & Hashcat 8900 & n=32768, r=24, p=1 & 202.4 ms & 35378.8 ms ($\approx$ 35 s) \\ \hline
Argon2id & John (argon2) & m=4096, t=150, p=1 & 199.7 ms & $2.69\times10^5$ ms ($\approx$ 4.5 min) \\ \hline
MFCHF-HOTP6 & John (argon2) & m=4096, t=150, p=1 & 201.6 ms & $2.7\times10^{11}$ ms ($\approx$ 8.6 yr) \\ \hline
MFCHF-TOTP6 & John (argon2) & m=4096, t=125, p=1 & 199.6 ms & $2.3\times10^{11}$ ms ($\approx$ 7.2 yr) \\ \hline
MFCHF-OOBA6 & John (argon2) & m=4096, t=135, p=1 & 200.9 ms & $\approx\infty$ (\textgreater 10,000 yr) \\ \hline
MFCHF-HSHA1 & John (argon2) & m=4096, t=150, p=1 & 199.8 ms & $\approx\infty$ (\textgreater 10,000 yr) \\ \hline
\end{tabular}%
}
\caption{Average verification and crack time for a variety of hash functions including MFCHF.}
\label{tab:results}
\end{table}

\twocolumn

%% file: 940-Appendix-Algorithms.tex
\section{Algorithms}
\label{app:algs}

\begin{algorithm}[H]
\caption{MFCHF with YubiKey (mfchf-hmacsha1)}
\label{alg:mfchf_hmacsha1}
\begin{algorithmic}[1]
\Require $\mathsf{HS1}$ is HMAC-SHA1 per RFC 2014 \cite{rfc2104}
\Require $\mathsf{H}$ is a password hash function (e.g., argon2)
\Function{Setup}{$\mathsf{password}$}
    \State $\mathsf{hmackey} \gets \mathsf{Random}(0, 2^{160})$
    \State $\mathsf{challenge} \gets \mathsf{Random}(0, 2^{160})$
    \State $\mathsf{salt} \gets \mathsf{Random}(0, 2^{256})$
    \State $\mathsf{response} \gets \mathsf{HS1}(\mathsf{hmackey}, \mathsf{challenge})$
    \State $\mathsf{digest} \gets \mathsf{H}(\mathsf{password} \odot \mathsf{hmackey} \odot \mathsf{salt})$
    \State $\mathsf{paddedkey} \gets \mathsf{hmackey} \oplus \mathsf{response}$
    \State $\mathsf{hash} \gets \{\mathsf{paddedkey}, \mathsf{challenge}, \mathsf{salt}, \mathsf{digest}\}$
    \State \Return $\mathsf{hash}, \mathsf{hmackey}$
\EndFunction
\Function{Verify}{$\mathsf{password}, \mathsf{response}, \mathsf{hash}$}
    \State $\{\mathsf{paddedkey}, \mathsf{salt}, \mathsf{digest}\} \gets \mathsf{hash}$
    \State $\mathsf{hmackey} \gets \mathsf{paddedkey} \oplus \mathsf{response}$
    \State $\mathsf{expected} \gets \mathsf{H}(\mathsf{password} \odot \mathsf{hmackey} \odot \mathsf{salt})$
    \If{$\mathsf{digest} \neq \mathsf{expected}$}
        \State \Return $\mathit{reject}$
    \EndIf
    \State $\mathsf{challenge} \gets \mathsf{Random}(0, 2^{160})$
    \State $\mathsf{response} \gets \mathsf{HS1}(\mathsf{hmackey}, \mathsf{challenge})$
    \State $\mathsf{paddedkey} \gets \mathsf{hmackey} \oplus \mathsf{response}$
    \State $\mathsf{hash} \gets \{\mathsf{paddedkey}, \mathsf{challenge}, \mathsf{salt}, \mathsf{digest}\}$
    \State \Return $\mathit{accept}, \mathsf{hash}$
\EndFunction
\end{algorithmic}
\end{algorithm}

\begin{algorithm}[H]
\caption{MFCHF with HOTP (mfchf-hotp6)}
\label{alg:mfchf_hotp6}
\begin{algorithmic}[1]
\Require $\mathsf{HOTP}$ is HOTP per RFC 4226 \cite{rfc4226}
\Require $\mathsf{H_1}$ is a password hash function (e.g., argon2)
\Require $\mathsf{H_2}$ is a standard hash function (e.g., sha256)
\Function{Setup}{$\mathsf{password}$}
    \State $\mathsf{target} \gets \mathsf{Random}(0, 10^6)$
    \State $\mathsf{hotpkey} \gets \mathsf{Random}(0, 2^{256})$
    \State $\mathsf{salt} \gets \mathsf{Random}(0, 2^{256})$
    \State $\mathsf{counter} \gets 1$
    \State $\mathsf{firstotp} \gets \mathsf{HOTP}(\mathsf{hotpkey},\mathsf{counter})~\%~10^6$
    \State $\mathsf{offset} \gets (\mathsf{target} - \mathsf{firstotp})~\%~10^{6}$
    \State $\mathsf{pad} \gets \mathsf{H_1}(\mathsf{password} \odot \mathsf{target} \odot \mathsf{salt})$
    \State $\mathsf{paddedkey} \gets \mathsf{hotpkey} \oplus \mathsf{pad}$
    \State $\mathsf{digest} \gets \mathsf{H_2}(\mathsf{pad})$
    \State $\mathsf{hash} \gets \{\mathsf{counter}, \mathsf{offset}, \mathsf{paddedkey}, \mathsf{salt}, \mathsf{digest}\}$
    \State \Return $\mathsf{hash}, \mathsf{hotpkey}$
\EndFunction
\Function{Verify}{$\mathsf{password}, \mathsf{otp}, \mathsf{hash}$}
    \State $\{\mathsf{counter}, \mathsf{offset}, \mathsf{paddedkey}, \mathsf{salt}, \mathsf{digest}\} \gets \mathsf{hash}$
    \State $\mathsf{target} \gets (\mathsf{offset} + \mathsf{otp})~\%~10^6$
    \State $\mathsf{pad} \gets \mathsf{H_1}(\mathsf{password} \odot \mathsf{target} \odot \mathsf{salt})$
    \If{$\mathsf{H_2}(\mathsf{pad}) \neq \mathsf{digest}$}
        \State \Return $\mathit{reject}$
    \EndIf
    \State $\mathsf{counter} \gets \mathsf{counter} + 1$
    \State $\mathsf{hotpkey} \gets \mathsf{paddedkey} \oplus \mathsf{pad}$
    \State $\mathsf{nextotp} \gets \mathsf{HOTP}(\mathsf{hotpkey},\mathsf{counter})~\%~10^6$
    \State $\mathsf{offset} \gets (\mathsf{target} - \mathsf{nextotp})~\%~10^{6}$
    \State $\mathsf{hash} \gets \{\mathsf{counter}, \mathsf{offset}, \mathsf{paddedkey}, \mathsf{salt}, \mathsf{digest}\}$
    \State \Return $\mathit{accept}, \mathsf{hash}$
\EndFunction
\end{algorithmic}
\end{algorithm}

\begin{algorithm}[H]
\caption{MFCHF with TOTP (mfchf-totp6)}
\label{alg:mfchf_totp6}
\begin{algorithmic}[1]
\Require $\mathsf{HOTP}$ is HOTP per RFC 4226 \cite{rfc4226}
\Require $T,T_0,T_X$ are TOTP times per RFC 6238 \cite{rfc6238}
\Require $\mathsf{H_1}$ is a password hash function (e.g., argon2)
\Require $\mathsf{H_2}$ is a standard hash function (e.g., sha256)
\Function{Setup}{$\mathsf{password}, \mathsf{w}$}
    \State $\mathsf{target} \gets \mathsf{Random}(0, 10^6)$
    \State $\mathsf{totpkey} \gets \mathsf{Random}(0, 2^{256})$
    \State $\mathsf{salt} \gets \mathsf{Random}(0, 2^{256})$
    \State $\mathsf{counter} \gets \lfloor(T-T_0)/T_X\rfloor$
    \For{$\mathsf{j}~\textbf{in}~[0\ldots\mathsf{w}]$}
        \State $\mathsf{otp} \gets \mathsf{HOTP}(\mathsf{totpkey},\mathsf{counter}+\mathsf{j})~\%~10^{d}$
        \State $\mathsf{offsets}[\mathsf{j}] \gets (\mathsf{target} - \mathsf{otp})~\%~10^{d}$
    \EndFor
    \State $\mathsf{pad} \gets \mathsf{H_1}(\mathsf{password} \odot \mathsf{target} \odot \mathsf{salt})$
    \State $\mathsf{paddedkey} \gets \mathsf{totpkey} \oplus \mathsf{pad}$
    \State $\mathsf{digest} \gets \mathsf{H_2}(\mathsf{pad})$
    \State $\mathsf{hash} \gets \{\mathsf{counter}, \mathsf{offsets}, \mathsf{paddedkey}, \mathsf{salt}, \mathsf{digest}\}$
    \State \Return $\mathsf{hash}, \mathsf{totpkey}$
\EndFunction
\Function{Verify}{$\mathsf{password}, \mathsf{otp}, \mathsf{hash}$}
    \State $\{\mathsf{counter}, \mathsf{offsets}, \mathsf{paddedkey}, \mathsf{salt}, \mathsf{digest}\} \gets \mathsf{hash}$
    \State $\mathsf{index} \gets \lfloor(T-T_0)/T_X\rfloor - \mathsf{counter}$
    \State $\mathsf{target} \gets (\mathsf{offsets}[\mathsf{index}] + \mathsf{otp})~\%~10^6$
    \State $\mathsf{pad} \gets \mathsf{H_1}(\mathsf{password} \odot \mathsf{target} \odot \mathsf{salt})$
    \If{$\mathsf{H_2}(\mathsf{pad}) \neq \mathsf{digest}$} \Return $\mathit{reject}$
    \EndIf
    \State $\mathsf{counter} \gets \lfloor(T-T_0)/T_X\rfloor$
    \State $\mathsf{totpkey} \gets \mathsf{paddedkey} \oplus \mathsf{pad}$
    \For{$\mathsf{j}~\textbf{in}~[0\ldots\mathsf{w}]$}
        \State $\mathsf{otp} \gets \mathsf{HOTP}(\mathsf{totpkey},\mathsf{counter}+\mathsf{j})~\%~10^{d}$
        \State $\mathsf{offsets}[\mathsf{j}] \gets (\mathsf{target} - \mathsf{otp})~\%~10^{d}$
    \EndFor
    \State $\mathsf{hash} \gets \{\mathsf{counter}, \mathsf{offsets}, \mathsf{paddedkey}, \mathsf{salt}, \mathsf{digest}\}$
    \State \Return $\mathit{accept}, \mathsf{hash}$
\EndFunction
\end{algorithmic}
\end{algorithm}

\vspace{-0.8em}

\begin{algorithm}[H]
\caption{MFCHF with OOBA (mfchf-ooba6)}
\label{alg:mfchf_ooba6}
\begin{algorithmic}[1]
\Require $(\mathsf{Enc},\mathsf{Dec})$ is public-key encryption
\Require $\mathsf{H}$ is a password hash function (e.g., argon2)
\Function{Setup}{$\mathsf{password}$, $\mathsf{pk}$}
    \State $\mathsf{target} \gets \mathsf{Random}(0, 36^6)$
    \State $\mathsf{firstotp} \gets \mathsf{Random}(0, 36^6)$
    \State $\mathsf{offset} \gets (\mathsf{target} - \mathsf{firstotp})~\%~36^{6}$
    \State $\mathsf{salt} \gets \mathsf{Random}(0, 2^{256})$
    \State $\mathsf{digest} \gets \mathsf{H}(\mathsf{password} \odot \mathsf{target} \odot \mathsf{salt})$
    \State $\mathsf{ct} \gets \mathsf{Enc}(\mathsf{firstotp}, \mathsf{pk})$
    \State $\mathsf{hash} \gets \{\mathsf{ct}, \mathsf{pk}, \mathsf{offset}, \mathsf{salt}, \mathsf{digest}\}$
    \State \Return $\mathsf{hash}$
\EndFunction
\Function{Verify}{$\mathsf{password}, \mathsf{otp}, \mathsf{hash}$}
    \State $\{\mathsf{pk}, \mathsf{offset}, \mathsf{salt}, \mathsf{digest}\} \gets \mathsf{hash}$
    \State $\mathsf{target} \gets (\mathsf{offset} + \mathsf{otp})~\%~10^6$
    \State $\mathsf{expected} \gets \mathsf{H}(\mathsf{password} \odot \mathsf{target} \odot \mathsf{salt})$
    \If{$\mathsf{digest} \neq \mathsf{expected}$} \Return $\mathit{reject}$ \EndIf
    \State $\mathsf{nextotp} \gets \mathsf{Random}(0, 36^6)$
    \State $\mathsf{offset} \gets (\mathsf{target} - \mathsf{nextotp})~\%~36^{6}$
    \State $\mathsf{ct} \gets \mathsf{Enc}(\mathsf{nextotp}, \mathsf{pk})$
    \State $\mathsf{hash} \gets \{\mathsf{ct}, \mathsf{pk}, \mathsf{offset}, \mathsf{salt}, \mathsf{digest}\}$
    \State \Return $\mathit{accept}, \mathsf{hash}$
\EndFunction
\end{algorithmic}
\end{algorithm}